\documentclass[12pt]{article}
\usepackage{latexsym}
\usepackage{amsmath,,calrsfs}
\usepackage{amsfonts}
\usepackage{amssymb}
\usepackage{amscd}
\usepackage{fancybox}
\usepackage{cite}
\usepackage{amsmath,amsfonts,amsbsy}
\usepackage{pstricks,pst-node}
\usepackage[small,bf,hang]{caption2}
\usepackage{slashed}
\usepackage{graphicx}

\usepackage{float}

\psset{unit=1.3cm,linewidth=.5pt,radius=.2}  % controls the size of diagrams

\usepackage{enumitem}                       % fancier lists
\usepackage{multirow}                     % tables cells spanning multiple rows
\usepackage{float}                          % to force floaters to stay Here
\usepackage{lscape}                         % landscape laid out pages
\usepackage{bm}

\def\hybrid{\topmargin -20pt    \oddsidemargin 0pt
        \headheight 0pt \headsep 0pt
        \textwidth 6.25in       % A4 paper
        \textheight 9.5in       % A4 paper
        \marginparwidth .875in
        \parskip 5pt plus 1pt   \jot = 1.5ex}

\hybrid

\newcommand{\beq}{\begin{equation}}
\newcommand{\eeq}{\end{equation}}
\newcommand{\bi}{\begin{itemize}}
\newcommand{\ei}{\end{itemize}}
\newcommand{\bt}{\begin{tabular}}
\newcommand{\et}{\end{tabular}}
\newcommand{\bc}{\begin{center}}
\newcommand{\ec}{\end{center}}

\newcommand{\ft}[2]{{\textstyle {\frac{#1}{#2}} }}

\newcommand{\be}{\begin{equation}}
\newcommand{\ee}{\end{equation}}
\newcommand{\bea}{\begin{eqnarray}}
\newcommand{\eea}{\end{eqnarray}}
\newcommand{\ba}{\begin{array}}
\newcommand{\ea}{\end{array}}

\def\bbox{{\,\lower0.9pt\vbox{\hrule \hbox{\vrule height 0.2 cm
\hskip 0.2 cm \vrule height 0.2 cm}\hrule}\,}}
\newcommand{\dsl}{\pa \kern-0.5em /}

\makeatletter \@addtoreset{equation}{section} \makeatother

\def\slashchar#1{\setbox0=\hbox{$#1$}           % set a box for #1
   \dimen0=\wd0                                 % and get its size
   \setbox1=\hbox{/} \dimen1=\wd1               % get size of /
   \ifdim\dimen0>\dimen1                        % #1 is bigger
      \rlap{\hbox to \dimen0{\hfil/\hfil}}      % so center / in box
      #1                                        % and print #1
   \else                                        % / is bigger
      \rlap{\hbox to \dimen1{\hfil$#1$\hfil}}   % so center #1
      /                                         % and print /
   \fi}

\def\eq#1{(\ref{#1})}
\newcommand{\w}[1]{\\[0.#1cm]}

\begin{document}

\begin{titlepage}%1
\begin{center}

\hfill UG-09-28 \\
\hfill DAMTP-2009-50\\
\hfill MIFP-09-32

\vskip 2cm

{\Large \bf  Massive 3D Supergravity}

\vskip 1.5cm

{\bf Roel Andringa\,$^1$, Eric A.~Bergshoeff\,$^1$, Mees de
Roo\,$^1$, Olaf Hohm\,$^1$,\\[0.5ex] Ergin Sezgin\,$^2$
and Paul K.~Townsend\,$^3$} \\

\vskip 30pt

{\em $^1$ \hskip -.1truecm Centre for Theoretical Physics, University of Groningen, \\
Nijenborgh 4, 9747 AG Groningen, The Netherlands \vskip 5pt }

{email: {\tt E.A.Bergshoeff@rug.nl, O.Hohm@rug.nl}} \\

\vskip 15pt

{\em $^2$ \hskip -.1truecm George and Cynthia Woods Mitchell
Institute for Fundamental Physics and Astronomy, Texas A\& M
University, College Station, TX 77843, USA}

{email: {\tt sezgin@tamu.edu}}

\vskip 15pt

{\em $^3$ \hskip -.1truecm Department of Applied Mathematics and
Theoretical Physics,\\
Centre for Mathematical Sciences, University of Cambridge,\\
Wilberforce Road, Cambridge, CB3 0WA, U.K. \vskip 5pt }

{email: {\tt P.K.Townsend@damtp.cam.ac.uk}} \\

\end{center}

\vskip 1cm

\begin{center} {\bf ABSTRACT}\\[3ex]
\end{center}

We construct the ${\cal N}=1$ three-dimensional supergravity  theory
with cosmological, Einstein-Hilbert, Lorentz Chern-Simons, and
general curvature squared terms. We determine the general
supersymmetric configuration, and find a family of  supersymmetric
adS vacua with the supersymmetric Minkowski vacuum as a limiting
case. Linearizing about the Minkowski vacuum, we find three classes
of unitary theories; one is  the supersymmetric extension of the
recently discovered `massive 3D gravity'.  Another is a  `new
topologically massive supergravity'  (with no Einstein-Hilbert term)
that  propagates a single $(2,\frac{3}{2})$ helicity supermultiplet.

\begin{minipage}{13cm}
\small

\end{minipage}

%\today

%\end{center}

%\noindent

\vfill

%\July 2008

\end{titlepage}

\newpage
\setcounter{page}{1}
\tableofcontents

\newpage

%%%%%%%%%%%%%%%%%%%%%%%%%%%%%%%%%%%%
%%%%%%%%%%%%%%%%%%%%%%%%%%%%%%%%%%%%%%%%%%%%%%%%%%%%%%%%%

\section{Introduction}\setcounter{equation}{0}

%%%%%%%%%%%%%%%%%%%%%%%%%%%%%%%%%%%%%%%%%%%%%%%%%%%%%%%%%

Gravitational theories in which the Einstein-Hilbert (EH) term is supplemented with curvature squared terms generically propagate
negative energy (ghost) modes in addition to the desired spin 2 graviton modes. An exception is the `$(R+R^2)$'  theory,  where $R$ is the curvature scalar,  because this can be shown to be equivalent to a scalar field coupled to gravity, with a potential that provides a mass for the associated spin zero particle in the Minkowski vacuum \cite{Bicknell}; we shall refer to this as ``scalar massive gravity'' (SMG). Recently, three of us showed that there is another exception in three spacetime dimensions \cite{Bergshoeff:2009hq}:  ghosts are
avoided if  (i) the EH term appears with the `wrong-sign' and (ii) the curvature-squared scalar is
\be\label{defK}
K=  R_{\mu\nu}R^{\mu\nu} - \frac{3}{8}R^2\, ,
\ee
where $R_{\mu\nu}$ is the Ricci tensor.  In its Minkowski vacuum, this ``new massive gravity'' (NMG) model propagates, unitarily,
two massive modes of helicities $\pm2$. Unitarity has since been confirmed in \cite{Nakasone:2009bn,Deser:2009hb}.  A more general model obtained by adding a Lorentz Chern-Simons (LCS) term was also considered in \cite{Bergshoeff:2009hq}, and this model propagates the two spin 2 modes with different masses; by taking one of the two masses
to infinity one gets the ``topologically-massive gravity'' (TMG) of \cite{Deser:1981wh}.

This  ``general massive gravity'' (GMG) model has an obvious
extension to include a cosmological constant, and this
`cosmological' GMG was investigated briefly in
\cite{Bergshoeff:2009hq}, allowing for either sign of the EH term as
in studies of cosmological TMG
\cite{Li:2008dq,Carlip:2008jk,Grumiller:2008qz,Maloney:2009ck}. In a
subsequent work \cite{Bergshoeff:2009aq}  the issue of unitarity and
stability in de Sitter (dS) and anti de Sitter (adS) vacua was
considered in detail for the `cosmological' NMG model. Other
investigations  of this model include \cite{others}. Of particular
interest are adS vacua since these could be associated with
potentially novel 2D conformal field theories (CFTs) on the adS
boundary. A major result of \cite{Bergshoeff:2009aq} was the finding
that  the central charge of this boundary CFT field theory is
negative whenever the bulk theory is unitary, and vice-versa, with
the exception of one case in which the bulk gravitons are absent and
the central charge vanishes. Essentially the same difficulty arises
in cosmological TMG; in this context, the `chiral gravity' program
initiated in \cite{Li:2008dq} may yield a resolution but this
remains unclear. One motivation for the study of 3D supergravity
models with curvature-squared terms is that supersymmetric adS vacua
may be `better behaved'  than generic adS vacua.

The ${\cal N}=1$ supergravity extension of the NMG model was already
considered briefly in  \cite{Bergshoeff:2009hq}, as was  the more
general model  with generic curvature-squared  term of the form
$(aK+ bR^2)$. The off shell supermultiplet containing the metric
(actually dreibein) and gravitino field  also contains an
`auxiliary' field $S$ \cite{Uematsu:1984zy} which really is
auxiliary when $b=0$, in the sense that its equation of motion is
algebraic. As noted in  \cite{Bergshoeff:2009hq}, the fully
non-linear supergravity theory must contain either an $S^4$ or an
$S^2R$ term, or both,  and the (non-zero) constant value of $S$ in
any adS vacuum depends on the coefficients of these terms. Thus, we
expect a cubic equation for $S$ with $R$-dependent coefficients. It
is not difficult to see that $S=0$ is necessarily a solution in the
absence of a cosmological term in the action, and this is sufficient
to deduce that the Minkowski vacuum of the GMG model is
supersymmetric. In fact, the Minkowski vacuum with $S=0$ is
supersymmetric quite generally, so linearization about this vacuum
of any of  the `$(aK+ bR^2)$' models  must yield a  theory in which
all modes (particles in the quantum theory) form  supermultiplets.
Any massive particles must have a definite helicity,  and it was
shown in \cite{Bergshoeff:2009hq} (by adapting earlier results of
\cite{Nishino:2006cb}) that  super-GMG propagates one supermultiplet
of  helicities $(2,\frac{3}{2})$ and another  of helicities
$(-2,-\frac{3}{2})$, generically with different masses.

In this paper we construct, in detail,  the off-shell supersymmetric
${\cal N}=1$ 3D supergravity  model with both generic
curvature-squared terms and cosmological constant.  To be specific,
we construct the 3D supergravity theory with action of the form
\be\label{basicact} I \ = \ \frac{1}{\kappa^2} \int \! d^3x \left\{e
\left[M L_C + \sigma L_{EH} + \frac{1}{m^2} L_K + \frac{1}{8\tilde
m^2} L_{R^2}\right] + \frac{1}{\mu}{\cal L}_{top}\right\}\;, \ee
where $e$ is the volume scalar density, $(M,m,\tilde m,\mu)$ are
mass parameters, $\sigma$ is a dimensionless parameter,  and
$\kappa$ is the 3D gravitational coupling required to ensure that
$I$ has dimensions of an action.  The individual Lagrangians in this
action are separately supersymmetric and they take the form
\begin{eqnarray}\label{listact}
L_C &=&  S + \ {\rm fermions}\;, \nonumber \\
L_{EH} &=& R - 2S^2  + \ {\rm fermions}\;, \nonumber \\
L_{K} &=& K -\frac{1}{2} S^2R - \frac{3}{2} S^4  + \ {\rm fermions}\;, \\
L_{R^2} &=& -16\left[ \left(\partial S\right)^2 - \frac{9}{4}\left(S^2+\frac{1}{6}R\right)^2\right]
+ \ {\rm fermions}\;, \nonumber \\
 {\cal L}_{top} &=& -\frac14 \varepsilon^{\mu\nu\rho} \left( R_{\mu\nu}{}^{ab}(\omega) \omega_{\rho ab}
+\frac23 \omega_{\mu}{}^{a}{}_{b} \omega_{\nu}{}^{b}{}_c
\omega_{\rho}{}^{c}{}_{a}\right) + \ {\rm fermions}\;, \nonumber
 \end{eqnarray}
where the `fermions'  provide the ${\cal N}=1$ supersymmetric
completion, and $\omega$ is the usual spin connection.

Note that $S$ does not appear in ${\cal L}_{top}$, which is the supersymmetric extension of the Lorentz Chern-Simons (LCS) term
\cite{Deser:1982sw,Deser:1982sv,vanNieuwenhuizen:1985cx}; this is because of the superconformal invariance of this term.  Note also that $S$ is indeed auxiliary as long as the $L_{R^2}$ term is absent, which is achieved by taking the limit $\tilde m^2\to\infty$. As summarized above, this limit yields a unitary theory of massive gravitons in the Minkowski vacuum with $S=0$ (which is a solution when $M=0$) but the presence of an  $S^2R$ term suggests that the ``effective'' balance of the two possible curvature-squared terms in a given (a)dS vacuum could depend on the (constant) value of $S$ in this vacuum, and this possibility motivates consideration of the generic theory. We do not know, a priori, which (if any) combination of   curvature-squared terms will allow a unitary theory in a non-Minkowski vacuum.

An important aspect of our construction is the full dependence of the bosonic action on the `auxiliary' field $S$, as given above.  This is crucial both for a classification of the possible maximally symmetric vacua, and for a determination of whether a given adS vacuum is supersymmetric (since one needs to know the value of $S$ in it).  In the absence of the curvature squared terms, i.e, in the limit that $m^2\to\infty$ and $\tilde m^2\to\infty$,  the $S$ field may be trivially eliminated and the resulting action then has a cosmological constant $\Lambda$ proportional to $M^2$.  Otherwise, the relation between
$\Lambda$ and $M^2$ is more complicated.  In fact, the possible maximally-symmetric vacua correspond to points on two curves in the $(\Lambda,M^2)$ plane.  All supersymmetric vacua lie on one of these two curves (actually a half-line) which (remarkably) is the same for all $\tilde m$; in other words, the presence or absence of the $L_{R^2}$ term in the action has no effect  on supersymmetric vacua, although we expect that it will affect the fluctuations and hence the analysis of unitarity/stability.  The endpoint  of the `supersymmetric curve' in the $(\Lambda,M^2)$ plane is the origin, corresponding to the supersymmetric Minkowski vacuum with $S=0$. Apart from this one special vacuum,  there is no obvious way to compare results with  those found in \cite{Bergshoeff:2009aq} for the purely bosonic theory (without the $S$ field). This unusual feature is due to the non-linearity of the $S$ equation of motion.  It means that the unitarity/stability analysis of \cite{Bergshoeff:20
 09aq}  must be undertaken anew,  but this  is encouraging because there is therefore the possibility of an improved outcome in regard to boundary CFT central charges.

The supersymmetric Minkowski and adS vacua are special cases of
supersymmetric solutions of the field equations. Because we have an
off-shell supersymmetry (in the sense that the supersymmetry
transformations close without the need to invoke field equations)
one can separate the question of whether a given field configuration
is  supersymmetric from the question of whether it solves any
particular set of field equations. Here we present a complete
analysis of the possible supersymmetric configurations, generalizing
an analysis presented in \cite{Gibbons:2008vi} for what amounts to
the special case in which the scalar field $S$ is constant and
non-zero. In this special case, we recover the generic
supersymmetric pp-wave configurations  that generalize the adS
vacua.  Particular subcases are known to solve the field equations of conformal 3D gravity \cite{Deser:2004wd},
TMG  \cite{AyonBeato:2004fq,Gibbons:2008vi}, and  NMG
\cite{AyonBeato:2009yq}; here we find the supersymmetric pp-wave
solutions of the generic 3D supergravity theory of the type under
consideration. We will comment on some features that can be read off
from these solutions as, for instance, critical values of the adS
length $\ell$ at which the pp-wave solution becomes locally
diffeomorphic to adS, indicating a generalized notion of `chiral
gravity'.

In the last part of this paper  we present a classification of  all
supergravity theories of  $(aK+ bR^2)$ type that are unitary in a
Minkowski vacuum, together with a detailed analysis of their
fermionic sectors.  We begin with a `canonical' analysis along the
lines of \cite{Deser:2009hb}; this throws up three classes of
unitary theories, which are the supersymmetrizations of the
following three classes of bosonic models:
\begin{itemize}

\item General massive gravity (GMG).  As summarized above, this propagates two massive gravitons of helicities $\pm 2$,
generically with different masses. This includes TMG and NMG as
special cases.

\item Scalar massive gravity (SMG). This is the parity-preserving theory with $a=0$ and `right-sign' EH term, equivalent to a scalar field coupled
to gravity. To the best of our knowledge, its supersymmetrization
has not been considered previously.

\item `New Topologically Massive Gravity' or NTMG. This is a  model in which the EH term is omitted.  It involves the `new' scalar $K$ but otherwise turns out to propagate a single helicity $2$ mode, like TMG, hence the name. The massless limit yields the `pure-K' theory considered in \cite{Deser:2009hb}.

\end{itemize}
This analysis does not yield the helicity content of the massive modes, so we then reconsider each of these three classes of unitary theories using covariant methods. In particular, we present a new proof that the GMG model propagates, unitarily, two spin 2 modes, and we verify the supermultiplet content of its supersymmetric extension.  The novel feature of super-SMG is a third-order equation for a vector spinor field that propagates, unitarily, two spin 1/2 modes.
The NTMG theory is new, even as a purely bosonic theory, so we consider this in more detail; in particular, we show that the linearized theory propagates a single massive mode of  helicity 2, just like TMG, and this becomes a supermultiplet of helicities $(2,\frac{3}{2})$, or $(-2,-\frac{3}{2})$, in the supersymmetric case.

Finally, we consider the linearized ${\cal N}=2$ super-GMG model. This is an obvious first step in an investigation of ${\cal N}>1$ 3D massive supergravities. It is also of interest in that it unifies the new spin 2 models with well-known spin 1 models.

%%%%%%%%%%%%%%%%%%%%%%%%%%%%%%%%%%%%
%%%%%%%%%%%%%%%%%%%%%%%%%%%%%%%%%%%

\section{${\cal N}=1$ massive supergravity}
\setcounter{equation}{0}

In this section we are going to determine the full non-linear ${\cal
N}=1$ supersymmetric off-shell invariants corresponding to the
action (\ref{basicact}). First, we give the off-shell ${\cal N}=1$
supergravity multiplet together with the known invariants
corresponding to the Einstein-Hilbert action with cosmological
constant and the LCS term. Next, we determine the curvature-square
invariants.

Our conventions are as follows. The metric  signature is `mostly
plus'.  All fermions are two-component Majorana spinors. We may
choose the  Dirac matrices  $\gamma^a$ ($a=0,1,2$), which satisfy
the anticommutation relation $\{ \gamma^{a},\gamma^{b}\}
=2\eta^{ab}$, to be real $2\times 2$ matrices, in which case the
Majorana spinors are also real. The Ricci tensor is
$R_{\mu\nu}\equiv R_{\rho\mu}{}^{\rho}{}_{\nu}=
\partial_{\rho}\Gamma^{\rho}_{\mu\nu}+\cdots$.

\subsection{Off-shell ${\cal N}=1$ supergravity multiplet}

The ${\cal N}=1$ supergravity multiplet in 3D consists of the
dreibein $e_{\mu}{}^{a}$ and the gravitino $\psi_{\mu}$, neither of  which propagates any modes
in  `pure'  supergravity but both  will start
propagating once higher-derivative terms are added. Off-shell closure requires a
real scalar auxiliary field $S$. The supersymmetry transformation
rules are
\bea\label{susy} \delta e_\mu{}^a &=& \frac12 {\bar\epsilon}
\gamma^a \psi_\mu\ ,
\\
\delta \psi_\mu &=& D_\mu(\hat{\omega})\epsilon +\frac12 S
\gamma_\mu\epsilon \ ,
\\\label{susy3}
\delta S &=&  \frac14 {\bar\epsilon} \gamma^{\mu\nu}
\psi_{\mu\nu}(\hat{\omega}) -\frac14 {\bar\epsilon} \gamma^\mu
\psi_\mu S\ , \eea
where
 \begin{eqnarray}
  D_{\mu}(\omega)\epsilon = \partial_{\mu}\epsilon + \frac{1}{4}
  \omega_{\mu}{}^{ab}\gamma_{ab}\epsilon\,,\qquad
   \psi_{\mu\nu}(\omega) =  \frac{1}{2}\left(D_{\mu}(\omega)\psi_{\nu}- D_{\nu}(\omega)\psi_{\mu}\right)\;.
 \end{eqnarray}
 The spin connection $\hat\omega$ is the spin-connection with torsion determined by  the super-torsion
constraint
 \bea
    D_{\mu}(\hat{\omega}) e_{\nu}{}^a - D_{\nu}(\hat{\omega})e_{\mu}{}^a =
          \frac{1}{2}\bar\psi_{\mu}\gamma^a\psi_{\nu}\,.
 \eea
Its solution reads
 \bea
   \hat{\omega}_{\mu}{}_{ab}(e,\psi) &=&
    \tfrac{1}{2}(R_{\mu ab} - R_{ab \mu } + R_{b\mu a})\,,
 \\
  R_{\mu\nu}{}^{a} &=& 2\partial_{[\mu} e_{\nu]}{}^a - \frac{1}{2}\bar\psi_{\mu}\gamma^a\psi_{\nu} \,.
 \eea
 In the following we denote by $D_{\mu}$ the covariant derivative
with respect to the standard spin connection $\omega=\omega(e)$ with
vanishing torsion. Whenever another connection is used, this will be
explicitly indicated.

The Lagrangians corresponding to cosmological constant,
Einstein-Hilbert and Lorentz Chern-Simons term have been constructed
long ago, and they are given by \cite{Deser:1982sw,Deser:1982sv,Uematsu:1984zy,vanNieuwenhuizen:1985cx}
\begin{eqnarray}
{ L}_C &=& S +\frac18 {\bar\psi}_\mu
\gamma^{\mu\nu}\psi_\nu \  ,
\\
{ L}_{EH} &=&  R -{\bar\psi}_\mu \gamma^{\mu\nu\rho}
D_\nu(\hat{\omega})\psi_\rho -2S^2\  ,
\\
e^{-1} {\cal L}_{top} &=& e^{-1}\varepsilon^{\mu\nu\rho} \left(
R_{\mu\nu}{}^{ab}(\hat{\omega}) \hat{\omega}_{\rho ab} +\frac23 \hat{\omega}_\mu{}^{a}{}_{b}
\hat{\omega}_{\nu}{}^{b}{}_c \hat{\omega}_{\rho}{}^{c}{}_{a}\right)+2{\bar
 {\cal R}}^\mu \gamma_\nu\gamma_\mu {\cal R}^\nu\ ,
\end{eqnarray}
where we defined the dual of the gravitino curvature,
 \be {\cal R}^\mu = e^{-1}\varepsilon^{\mu\nu\rho}
D_\nu(\hat{\omega})\psi_\rho\;. \ee

\subsection{Yang-Mills multiplets and the Riemann invariant}

Here we are going to determine the ${\cal N}=1$ supersymmetric
extension of the square of the Riemann tensor. This can be done very
efficiently by introducing a torsionful spin connection which allows the problem to be
reduced to one of coupling Yang-Mills multiplets to supergravity.

The off-shell Yang-Mills multiplet consists of a vector field and a
Majorana spinor, both transforming in the adjoint representation of
some gauge group. We denote these fields by $A_{\mu}{}^I$ and $\chi^I$,
where $I$ is a Lie algebra index. The supersymmetry transformations
are \be\label{YMrule}
   \delta A_{\mu}{}^I = -\bar\epsilon\gamma_{\mu}\chi^I\,,\qquad
   \delta\chi^I = \tfrac{1}{8}\gamma^{\mu\nu}\hat{F}_{\mu\nu}{}^I\epsilon\,,
\ee with the super-covariant field strength \be
   \hat{F}_{\mu\nu}{}^I = \partial_{\mu} A_{\nu}{}^I - \partial_{\nu} A_{\mu}{}^I
     + f_{JL}{}^I A_{\mu}{}^J A_{\nu}{}^L+2\bar{\psi}_{[\mu}\gamma_{\nu]}\chi^{I}\,.
\ee The locally supersymmetric $F^2$ invariant reads
 \begin{eqnarray}\label{superYM}
  {\cal L}_{\rm SYM} &=& -\frac{1}{4}eF^{\mu\nu I}F_{\mu\nu}{}^{I}-2e\bar\chi^{I}
  \gamma^{\mu}
  (D_{\mu}\chi)^I+
  \frac{1}{2}eF_{\mu\nu}{}^{I}\bar{\psi}_{\rho}\gamma^{\mu\nu}\gamma^{\rho}\chi^{I}
  +eS\bar{\chi}^{I}\chi^{I} \\ \nonumber
  &&-\frac{1}{2}e\bar{\chi}^{I}\chi^{I}\bar{\psi}_{\mu}\psi^{\mu}+
  \frac{1}{8}e\bar{\chi}^{I}\chi^{I}\bar{\psi}_{\mu}\gamma^{\mu\nu}\psi_{\nu}\;.
 \end{eqnarray}
Let us note that here the covariant derivative acting on $\chi^{I}$
is the ordinary covariant derivative with respect to the torsionless
spin connection. Introducing the super-covariant spin connection
$\hat{\omega}$ would change the coefficient of the second four-fermi
term. However, the quartic fermion couplings cannot be fully
absorbed into the spin connection and so we keep the standard
covariant derivative.

The next step is to realize that the spin connection can be
redefined such that it transforms under supersymmetry precisely as a
Yang-Mills gauge potential. For this we use the auxilliary field $S$
to define a torsionful connection as follows
 \be
  \Omega_{\mu}^{\pm ab} \ = \ \hat{\omega}_{\mu}{}^{ab} \pm S
  \varepsilon_{\mu}{}^{ab}\;.
 \ee
The supersymmetry transformations on $\psi_{\mu}$ and $S$ can in
turn be rewritten as
 \bea\label{newsusy}
  \delta\psi_{\mu} \ = \  D_{\mu}(\Omega^-)\epsilon\,, \qquad
  \delta S \ = \
  \tfrac{1}{4}\bar{\epsilon}\gamma^{\mu\nu}\psi_{\mu\nu}(\Omega^{-})\;.
 \eea
Here we have introduced the gravitino curvature with respect to
$\Omega^{-}$, i.e., explicitly
 \bea
  \psi_{\mu\nu}(\Omega^{-}) \ = \ \tfrac{1}{2}\left(
  D_{\mu}(\Omega^{-})\psi_{\nu}-D_{\mu}(\Omega^{-})\psi_{\nu}\right)\;.
 \eea
While the original spin connection $\hat{\omega}(e,\psi)$ transforms
under supersymmetry as
 \bea
  \delta\hat{\omega}_{\mu ab} = -\tfrac{1}{2}
   \bar\epsilon\left(\gamma_{\mu}\psi_{ab}(\Omega^{-}) -
   2\gamma_{[a}\psi_{b]\mu}(\Omega^{-})\right)
   +\tfrac{1}{2}S\bar{\psi}_{\mu}\gamma_{ab}\epsilon\,,
 \eea
the supersymmetry rule for $\Omega^{+}$ simplifies to
 \bea
  \delta\Omega_{\mu ab}^{+} \ = \ -\bar{\epsilon}\gamma_{\mu}\psi_{ab}(\Omega^{-})\;.
 \eea
We observe that there is no mixing left between Lorentz and world
indices. Consequently, the supersymmetry rule coincides with the one
for the gauge potential (\ref{YMrule}) if we treat the Lorentz
indices as Yang-Mills indices and if we identify
$\psi^{ab}(\Omega^{-})$ with the fermionic partner. To prove that
$\left(\Omega_{\mu}^{+\;ab},\psi^{ab}(\Omega^{-})\right)$ transforms
as a Yang-Mills vector multiplet it remains to check the
supersymmetry variation of $\psi^{ab}(\Omega^{-})$. We first observe
that
 \bea
   \delta \psi_{ab}(\Omega^{-}) \ = \
  \tfrac{1}{8}e_{a}{}^{\mu}e_{b}{}^{\nu}\hat{R}_{\mu\nu
  cd}(\Omega^{-})\gamma^{cd}\epsilon\;.
 \eea
This is almost of the required form, except that the connection is
$\Omega^{-}$ instead of $\Omega^{+}$ and that the index pairs are in
the `wrong' order. However, due to the torsionful connection the
standard Bianchi identity no longer holds but rather we have
 \bea\label{exchange}
  \hat{R}_{ab\;cd}(\Omega^{+}) \ = \ \hat{R}_{cd\;ab}(\Omega^{-})\;,
 \eea
where we have introduced the super-covariant form of the Riemann
tensor,
 \bea
  \hat{R}_{\mu\nu}{}^{ab}(\Omega^{+}) \ = \
  R_{\mu\nu}{}^{ab}(\Omega^{+})+2\bar{\psi}_{[\mu}\gamma_{\nu]}\psi^{ab}(\Omega^{-})\;.
 \eea
The generalized Bianchi identity (\ref{exchange}) can be easily
derived by writing out the explicit $S$ dependence,
 \bea\label{RiemS}
  R_{\mu\nu}{}^{ab}(\Omega^{\pm}) \ = \ R_{\mu\nu}{}^{ab}(\hat{\omega})
  \pm\tfrac{1}{2}S\bar{\psi}_{\mu}\gamma^{ab}\psi_{\nu}\pm2\partial_{[\mu}S\;\varepsilon_{\nu]}{}^{ab}
  +2S^2e_{[\mu}{}^{a}e_{\nu]}{}^{b}\;.
 \eea
In total this implies that
$(\Omega_{\mu}^{+\;ab},\psi^{ab}(\Omega^{-}))$ transforms precisely
as a Yang-Mills vector multiplet.

Finally, we can give the supersymmetric extension of the square of
the Riemann tensor simply by specializing (\ref{superYM}) to the
multiplet $\left(\Omega_{\mu}^{+\;ab},\psi^{ab}(\Omega^{-})\right)$,
 \begin{eqnarray}
  {\cal L} &=&  -\frac{1}{4}eR^{\mu\nu ab}(\Omega^{+})R_{\mu\nu
  ab}(\Omega^{+})-2 e\bar{\psi}_{ab}(\Omega^{-})\gamma^{\mu}D_{\mu}\psi^{ab}(\Omega^{-})\\ \nonumber
  &&+\frac{1}{2}eR_{\mu\nu
  ab}(\Omega^{+})\bar{\psi}_{\rho}\gamma^{\mu\nu}\gamma^{\rho}\psi^{ab}(\Omega^{-})
  +eS\bar{\psi}_{ab}(\Omega^{-})\psi^{ab}(\Omega^{-}) \\ \nonumber
  &&-\frac{1}{2}e\bar{\psi}^{ab}(\Omega^-)\psi_{ab}(\Omega^-)\bar{\psi}_{\mu}\psi^{\mu}+
  \frac{1}{8}e\bar{\psi}^{ab}(\Omega^-)\psi_{ab}(\Omega^-)\psi_{\mu}\gamma^{\mu\nu}\psi_{\nu}\;.
 \end{eqnarray}
Here we stress again, that unless stated differently the covariant
derivative is with respect to $\omega(e)$. Since the Riemann tensor
is equivalent to the Ricci tensor in 3D this result amounts to
supersymmetrizing $R^{\mu\nu}R_{\mu\nu}$. Using
 \bea
  \varepsilon^{\mu\nu\rho}\varepsilon^{abc}R_{\nu\rho bc} \ = \
  4G^{\mu a}\;, \qquad
  R_{\mu\nu ab} \ = \ \varepsilon_{\mu\nu\rho}\varepsilon_{abc}G^{\rho c}\;,
 \eea
where $G_{\mu a}$ is the Einstein tensor, one finds for the bosonic
action
 \begin{eqnarray}\label{Riembos}
  e^{-1}{\cal L} &=&  -\left(
  R^{\mu\nu}R_{\mu\nu}-\tfrac{1}{4}R^2\right)+2\partial^{\mu}S\partial_{\mu}S-S^2R-3S^4\;.\\
  \nonumber
  %&&+\cdots\;,
 \end{eqnarray}

\subsection{Scalar multiplets and the Ricci scalar invariant}

After
having determined the supersymmetric extension of the square of the
Riemann tensor, and hence of the Ricci tensor, the only independent invariant left in
3D is the supersymmetrization of the square of the Ricci scalar $R$. This can be reduced to the
problem of coupling an off-shell scalar multiplet to supergravity, in a similar way that we reduced
the earlier problem to one of coupling a Yang-Mills multiplet to supergravity.

An off-shell ${\cal N}=1$ scalar multiplet in 3D consists of a real
scalar $\phi$, a Majorana fermion $\lambda$ and a real auxiliary
scalar $f$. Its Lagrangian, after coupling to supergravity, reads
 \begin{eqnarray}\label{scalaraction}
  {\cal L} &=&
  -eg^{\mu\nu}\partial_{\mu}\phi\partial_{\nu}\phi-\frac{1}{4}e\bar{\lambda}\gamma^{\mu}D_{\mu}\lambda
  +\frac{1}{16}ef^2 %\\ \nonumber
  +\frac{1}{8}eS\bar{\lambda}\lambda+\frac{1}{2}e\bar{\psi}_{\mu}\gamma^{\nu}\gamma^{\mu}\partial_{\nu}\phi
  \lambda \\ \nonumber
  &&+\frac{1}{32}e\bar{\lambda}\lambda\bar{\psi}_{\mu}\psi^{\mu}-\frac{1}{64}e\bar{\lambda}\lambda
  \bar{\psi}_{\mu}\gamma^{\mu\nu}\psi_{\nu}\;.
 \end{eqnarray}
The supersymmetry rules are
 \begin{eqnarray}\label{scalarsusy}
  \delta \phi &=& \frac{1}{4}\bar{\epsilon}\lambda \;, \\
  \label{scalarsusy2}
  \delta \lambda &=&
  \hat{\slashed{D}}\phi\;\epsilon-\frac{1}{4}f\epsilon\;, \\
  \label{scalarsusy3}
  \delta f &=&
  -\bar{\epsilon}\hat{\slashed{D}}\lambda+\frac{1}{2}S\bar{\epsilon}\lambda\;,
 \end{eqnarray}
where the super-covariant derivatives are given by
 \begin{eqnarray}
  \hat{D}_{\mu}\phi &=&
  \partial_{\mu}\phi-\frac{1}{4}\bar{\psi}_{\mu}\lambda\;, \\
  \hat{D}_{\mu}\lambda &=&
  D_{\mu}(\hat{\omega})\lambda-\hat{\slashed{D}}\phi\psi_{\mu}+\frac{1}{4}f\psi_{\mu}\;.
 \end{eqnarray}

We will now show that
 \bea
  (\phi,\lambda,f) \ \equiv \
  (S,\gamma^{\mu\nu}\psi_{\mu\nu}(\Omega^{-}),\hat{R}(\Omega^{\pm}))\;,
 \eea
where
 \bea
  \hat{R}(\Omega^{\pm}) \ = \
  R(\hat{\omega})+6S^2+2\bar{\psi}_{\mu}\gamma_{\nu}\psi^{\mu\nu}(\Omega^{-})
  +\frac{1}{2}S\bar{\psi}_{\mu}\gamma^{\mu\nu}\psi_{\nu}\,,
 \eea
transforms under local supersymmetry precisely as required by
(\ref{scalarsusy}), (\ref{scalarsusy2}) and (\ref{scalarsusy3}).
First, we infer from (\ref{newsusy}) that $S$ transforms as the
scalar component. Moreover, it is easily checked that
 \bea
  \delta \left(\gamma^{\mu\nu}\psi_{\mu\nu}(\Omega^{-})\right)  \ = \
  \hat{\slashed{D}}S\epsilon-\frac{1}{4}\hat{R}(\Omega^{\pm})\epsilon\;,
 \eea
i.e., the gamma trace of $\psi_{\mu\nu}(\Omega^{-})$ transforms as
the spinor component. It takes a little bit more work to check the
supersymmetry variation of $\hat{R}(\Omega^{\pm})$. Using
 \bea
  \gamma_{\nu}D_{\mu}\psi^{\mu\nu}(\Omega^{-}) \ = \
  \frac{1}{2}\slashed{D}(\gamma^{\mu\nu}\psi_{\mu\nu}(\Omega^{-}))
  -\frac{1}{2}\varepsilon^{\mu\nu\rho}
  D_{\mu}\psi_{\nu\rho}(\Omega^{-})\;,
 \eea
and
 \bea
  \varepsilon^{\mu\nu\rho}D_{\mu}\psi_{\nu\rho}(\Omega^{-}) \ =
  \
  \frac{1}{2}G^{\mu\nu}\gamma_{\mu}\psi_{\nu}-\frac{1}{2}\varepsilon^{\mu\nu\rho}
  D_{\mu}(S\psi_{\nu})\gamma_{\rho}\;,
 \eea
one may verify that
 \bea
  \delta \hat{R}(\Omega^{\pm}) \ = \
  -\bar{\epsilon}\slashed{\hat{D}}(\gamma^{\mu\nu}\psi_{\mu\nu}(\Omega^{-}))+\frac{1}{2}S\bar{\epsilon}
  \gamma^{\mu\nu}\psi_{\mu\nu}(\Omega^{-})\;,
 \eea
as required. Thus, we can use the supersymmetry of
(\ref{scalaraction}) to construct directly the $R^2$ invariant,
 \begin{eqnarray}
  {\cal L}_{R^2} &=&
  \frac{1}{16}e\hat{R}^2(\Omega^{+})+\frac{1}{4}e\bar{\psi}_{\mu\nu}(\Omega^{-})\gamma^{\mu\nu}
  \slashed{D}\gamma^{\rho\sigma}\psi_{\rho\sigma}(\Omega^{-})-e\partial^{\mu}S\partial_{\mu}S
  \\ \nonumber
  &&-\frac{1}{8}eS\bar{\psi}_{\mu\nu}(\Omega^{-})\gamma^{\mu\nu}\gamma^{\rho\sigma}\psi_{\rho\sigma}(\Omega^{-})
  +\frac{1}{2}e\bar{\psi}_{\mu}\gamma^{\nu}\gamma^{\mu}\partial_{\nu}S\gamma^{\rho\sigma}\psi_{\rho\sigma}(\Omega^{-})
  \\ \nonumber
  &&-\frac{1}{32}e\bar{\psi}_{\mu\nu}(\Omega^-)\gamma^{\mu\nu}\gamma^{\rho\sigma}\psi_{\rho\sigma}(\Omega^-)
  \bar{\psi}_{\lambda}\psi^{\lambda}+\frac{1}{64}e\bar{\psi}_{\mu\nu}(\Omega^-)\gamma^{\mu\nu}\gamma^{\rho\sigma}
  \psi_{\rho\sigma}(\Omega^-)
  \bar{\psi}_{\lambda}\gamma^{\lambda\tau}\psi_{\tau}\;.
 \end{eqnarray}
The leading terms corresponding to the first line were given in
\cite{Nishino:2006cb}. Its bosonic part reads explicitly
 \begin{eqnarray}\label{Rbos}
  {\cal L}_{R^2} &=&
  \frac{1}{16}eR^2-e\, \partial^{\mu}S\partial_{\mu}S+\frac{3}{4}eS^2R+\frac{9}{4}eS^4\;.
 \end{eqnarray}
In total we have determined the complete supersymmetrisation of the
bosonic actions given by (\ref{Riembos}) and (\ref{Rbos}) from which
the form (\ref{listact}) given in the introduction readily follows.

\section{Supersymmetric configurations}

Before proceeding to consider solutions of the field equations, we
shall first determine which  bosonic field configurations are
supersymmetric. By definition, these are configurations that  admit
a Killing spinor, defined as a non-zero solution for $\kappa$ to the
equation \be\label{KSE} \left(D_\mu + \frac{1}{2} \gamma_\mu
S\right)\kappa =0\, , \ee which is obtained by setting to zero the
supersymmetry variation of the gravitino field, specializing to
bosonic field confgurations and replacing the anticommuting spinor
$\epsilon$ by the commuting spinor field $\kappa$.  The $S$ term may
be viewed as a torsion part of the spin connection.  The
integrablity condition of the Killing spinor equation is
\be\label{susycon1} \left(G^{\mu\nu} - g^{\mu\nu} S^2 -
e^{-1}\varepsilon^{\mu\nu\rho}\partial_\rho S\right)\gamma_\nu
\kappa =0\, . \ee It follows from this equation that the only
maximally supersymmetric field configurations are Minkowski space, with $S=0$,  and anti-de Sitter space,
with $G_{\mu\nu}= S^2g_{\mu\nu}$
for constant non-zero $S$, so the main interest in what follows will
be in other configurations that preserve 1/2 supersymmetry.

To begin with, we may easily deduce some other relations from
(\ref{susycon1}). By contracting with $\gamma_\mu$ one finds that
\be\label{susycon2}
 \gamma^\mu \kappa\,  \partial_\mu S =
\frac{3}{2}\left(S^2 + \frac{1}{6}R\right) \kappa\, , \ee which in
turn implies that \be\label{gensusycon} \left(\partial S\right)^2 =
\frac{9}{4} \left( S^2 + \frac{1}{6}R\right)^2\, . \ee Remarkably,
this is equivalent to the vanishing of  the bosonic part of
$L_{R^2}$. This does not mean that the $L_{R^2}$ term is irrelevant
to the field equations because its variation could still be
non-zero. In the case of  maximally symmetric supersymmetric vacua,
for which  $S$ is constant, even the variation of $L_{R^2}$ is zero,
so the possibilities for such vacua are unaffected by the presence
of the $L_{R^2}$ term. Moreover, all contributions of the curvature
squared terms to the field equations, including those of $K$, vanish
when evaluated for maximally symmetric supersymmetric
configurations, as we will show in sec.~\ref{maxsusy}. However,
these contributions could affect other non-supersymmetric vacua, and
supersymmetric non-vacuum solutions. Also the second variation, of
relevance to perturbative unitarity and stability, is generically
non-vanishing.

\subsection{The null Killing vector field}

To make further progress, we observe that the existence of a Killing spinor
implies the existence of a null vector field:
\be
V^\mu = \bar\kappa \gamma^\mu \kappa\, , \qquad V^2=0\, .
\ee
Note that since $\bar\kappa \kappa \equiv 0$, a direct consequence of  (\ref{susycon2}) is
the relation
\be\label{invS}
V^\mu \partial_\mu S=0\, .
\ee In other
words, $S$ is constant on orbits of  $V$. Similarly, an immediate consequence of
(\ref{susycon1}) is the relation
\be\label{susyV}
\left(G^{\mu\nu} - g^{\mu\nu} S^2 - e^{-1}\varepsilon^{\mu\nu\rho}\partial_\rho S\right)V_\nu =0\, ,
\ee
which implies, in particular,  that  $G^{\mu\nu} V_{\mu}V_\nu=0$.

The vector field $V$ is  covariantly constant with respect to the connection
with torsion defined by the Killing spinor equation. Explicitly,
this condition reads\footnote{This relation was
previously derived in  \cite{Gibbons:2008vi} under the assumption of
constant $S$.}
\be\label{DVS}
e\, D^\mu  V^\nu  =  S\, \varepsilon^{\mu\nu\rho}
V_\rho \, .
\ee
This implies that $D_{(\mu}V_{\nu)}=0$ and hence that
$V$ is a Killing vector field (KVF). It also implies that
\be\label{extrasusy}
\varepsilon^{\mu\nu\rho} \partial_\nu V_\rho = -2e S V^\mu\, .
\ee

\subsection{Adapted coordinates}

The full implications of (\ref{susyV}) and (\ref{extrasusy})  can be
analysed by choosing coordinates that are adapted to the null KVF.
The general 3-metric with null Killing vector $V=\partial_v$ takes
the form \be g = h_{ij} dx^i dx^j + 2A_i dx^i dv \qquad
\left(i,j=1,2\right)\;, \ee where the (not necessarily invertible)
symmetric 2-tensor field $h_{ij}$ and the 1-form $A_idx^i$ are
independent of $v$.  We may choose new coordinates $x^{i}=(u,x)$
such that \be dx = \sqrt{h_{22}}\, dx^2 + F dx^1\, , \qquad A_i dx^i
= f(x,u)\, du \;, \ee for some positive function $f$, and function
$F$ such that $\partial_2 F=\partial_1 \sqrt{h_{22}}$. We may then
shift $v$ by a function of $x$ and $u$ so as to remove the $dudx$
term in the metric. We thus arrive at a metric of  the form \be g =
dx^2 + 2 f(x, u) du dv + h(x, u) du^2 \;,  \label{s1} \ee where
$f(x,u)$ is
everywhere positive. %The inverse is given by \be g^{-1} =
%\partial_x
%\partial_x + \frac{2}{f} \partial_u\partial_v - \frac{h}{f^2}
%\partial_v\partial_v\, . \ee
For this metric we have \be \sqrt{|g|}= f\, ,
\qquad V_u=f\;, \qquad V_v=V_x=0\, . \ee

We are now in a position to analyse the full content of
(\ref{extrasusy}). The $u$-component is an identity. The
$x$-component tells us that $\partial_v f=0$, which we already know.
The $v$ component involves a choice of sign for $\varepsilon^{xuv}$,
which amounts to a choice of one of the two irreducible
representations of the Clifford algebra spanned by the 3D Dirac
matrices and their products. For the choice \be\label{sign}
\varepsilon^{xuv} =
 1 \ee we find that\footnote{The sign of $S$ differs for the other choice, such that
the restrictions on the metric implied by supersymmetry become  independent of the choice
of Dirac matrices.}
\be\label{Sresult} S= - \partial_x \log \sqrt{f}\, . \ee A
computation of the  Ricci tensor yields the result
\begin{eqnarray}\label{riccicom}
R_{\mu\nu} dx^\mu dx^\nu &=& -2\left(S^2  -  \partial_x S\right)dx^2 -2f\left(2S^2 -  \partial_x S\right) dudv + 2\partial_u S dxdu \nonumber \\
&& - \ \left( S\partial_x h +2hS^2 + \frac{1}{2} \partial_x^2 h\right) du^2\, ,
\end{eqnarray}
where we have used (\ref{Sresult}).  This gives the Ricci scalar
\be\label{s3} R= -6S^2 + 4 \partial_x S\, , \ee in agreement  with
(\ref{gensusycon}).
%The Einstein tensor is
%\be G = S^2 g -
%2f\partial_x S dudv + 2 \partial_u S dx du - \left[ 2h
%\partial_x S + S\partial_x  h + \frac{1}{2}\partial_x^2 h\right]
%du^2\, , \ee which yields
We then find  that \be\label{Gup} \left(G - S^2
g\right)^{\mu\nu}\partial_\mu\partial_\nu  = 2 \frac{\partial_u
S}{f}
\partial_x\partial_v -  2 \frac{\partial_x S}{f} \partial_u
\partial_v - \frac{1}{f^2} \left(S\partial_x h + \frac{1}{2}
\partial_x^2 h \right) \partial_v\partial_v\;.
\ee We can now use this in  the  integrability condition
(\ref{susyV}). The $u$ and $v$ components are identities. The
$x$-component implies that $\partial_v S=0$, in agreement with
(\ref{invS}).

We have now established that a bosonic configuration of 3D
supergravity  is supersymmetric if it takes the form
\be\label{susymetric} ds^2 = dx^2 + 2f(u,x) dudv + h(u,x) du^2\, ,
\qquad S= - \partial_x \log \sqrt{f}\, , \ee where the functions $f$
and $h$ are arbitrary, except that $f$ is everywhere positive, and
the sign of $S$ depends on the choice of Dirac matrices.

\subsection{Constant $S$}

Let us now spell out the condition (\ref{susymetric}) for the case
that $S$ is constant. If we set $S=\pm 1/\ell$, for finite constant
$\ell$, then $f(u,x)=A(u)\exp\left(\mp 2x/\ell\right)$ for some
function $A(u)$, which we may set to unity without loss of
generality; we then have the metric
\be\label{ppmetric} ds^2 = dx^2 + e^{\mp 2x/\ell}  dudv +
h(u,x) du^2\;. \ee
This has the general form of a pp-wave metric; the special case of
$h\equiv 0$ yields a metric that is  locally isometric to adS, for
either choice of sign. Each choice  yields a chart that extends to a
horizon (at $x\to \pm \infty$) that separates the two charts. Taken
together,  the two charts cover the whole of adS except for the
horizon, although the sign of $S$ changes across the horizon. Thus,
it is really $S^2$ that is constant in the adS vacuum, rather than
$S$.  In the  limiting case that $\ell\to \infty$ (i.e. $S\to 0$) we
find the metric \be ds^2 = dx^2 + 2 dudv + h(u,x) du^2\,  , \ee
which is the pp-wave in a Minkowski background.

Here we shall find the Killing spinor admitted by the general (adS) pp-wave configuration.
Starting from the metric \eq{ppmetric} with lower sign in the exponent for concreteness,
setting  $\ell=1$ for notational simplicity, and changing coordinates as $e^x=r$, the metric takes
the form
\be ds^2= \frac{dr^2}{r^2} + 2r^2 du dv + h(u,r) du^2\ , \label{pp}
\ee
where $h(u,r)$ is an undetermined function. Next, we  choose the basis 1-forms as
\footnote{The $\pm$ labels denote flat indices. To be specific, given a vector
$v_{a}$ in the tangent space, we define the light-cone indices in a local Lorentz
frame as $v_{\pm}=\tfrac{1}{\sqrt{2}}(\pm v_0+v_1)$.}
\be
e^+ = r dv +\frac{h}{2r} du\ ,\qquad e^- = r du\ ,\qquad e^2=\frac{1}{r} dr\ .
\ee
It follows that the only non-vanishing components of the spin connection one-form are
\be
\omega^{+2}=rdv+ r\partial_r \left(\frac{1}{2r}\right) du\ ,\qquad \omega^{-2}=rdu\ .
\label{scc}
\ee
The Killing spinor equation $ (d + \frac14 \omega^{ab}\gamma_{ab} -
\frac12 e^a\gamma_a) \kappa=0$  takes the form
\be d\kappa + \frac12 \left( \omega^{+2}\gamma_{+2} +
\omega^{-2}\gamma_{-2}\right)\kappa -\frac12 \left(e^+
\gamma_+ + e^- \gamma_- + e^2 \gamma_2\right) \kappa =0\ . \ee
A convenient choice of $\gamma$ matrices is
\be
\gamma_0=i\sigma_2\ ,\qquad \gamma_1=\sigma_1\ ,\qquad \gamma_2= \sigma_3\ .
\ee
Writing the spinor parameter as
\be \kappa= \left(
  \begin{array}{c}
    \psi \\
    \chi \\
  \end{array}
\right)\;, \ee
we find that
\bea d\psi &=& {\sqrt 2}r\chi\,dv + \frac1{\sqrt
2} \chi(1+r\partial_r )\left(\frac{h}{2r}\right) du
+\frac{1}{2r} \psi dr\ ,
\label{k1}\\[0.5pt]
d\chi &=&  -\frac{1}{2r} \chi dr\ .
\eea
%
%Recalling that $n\ne 0,2$ in (\ref{pp}),
The solution to these equations is given by
\be \psi= \psi_0 {\sqrt r}\ ,\qquad \chi=0\;, \ee
where $\psi_0$ is an arbitrary constant. This means that half of
supersymmetry is broken, in the sense that we have a Killing spinor
$\kappa_0$ given by\footnote{Note that this result is considerably
simpler in form than that found in \cite{Gibbons:2008vi} due to our
different choice of basis one-forms.}
\be \kappa_0 = {\sqrt r}\eta_-\ ,
\label{ks1}
\ee
where $\eta_-$ is a {\it single} Majorana-Weyl spinor in $1+1$
dimensions satisfying $\gamma_2\eta_-=-\eta_-$. Nota also that since
$\chi=0$, the term containing the function $h(u,r)$ in \eq{k1} drops
out, and consequently the Killing spinor \eq{ks1} exists for a {\it
generic} pp-wave solution, not depending on the detailed form of
$h(r,u)$.

If we specialize to the $adS_3$ metric, which amounts to setting
$h=0$, the solution is given by
\be
\psi = {\sqrt r} (\psi_0 + {\sqrt 2}  v\chi_0)\ ,\qquad \chi=\frac{\chi_0}{\sqrt r}\ ,
\ee
where $\psi_0$ and $\chi_0$ are arbitrary constants. As expected, this means a symmetry enhancement, since the Killing spinor now takes the form \cite{Gibbons:2008vi}
\be
\kappa_0 = r^{-1/2} \eta_- + r^{1/2} \left( \eta_+ + v\gamma_+ \eta_-\right) ,
\ee
where $\eta_\pm$ are constant spinors satisfying $\gamma_2
\eta_\pm=\pm \eta_\pm$, and $\kappa_0$ now decomposes into {\it two}
independent Majorana-Weyl spinors from the $1+1$ dimensional point
of view.

%%%%%%%%%%%%%%%%%%%%%%%%%%%%

\section{Field equations and solutions }
\setcounter{equation}{0}

{}From (\ref{basicact}) and (\ref{listact}) we see that the bosonic action of the generic 3D supergravity theory of interest is
\begin{eqnarray}\label{bosonicaction}
I  &=&  \frac{1}{\kappa^2} \int \!
d^3x \left\{e \left[M S + \sigma \left( R-2S^2 \right)+ \frac{1}{m^2} \left(K - \frac{1}{2} S^2 R - \frac{3}{2}S^4\right)\right]  \right.\nonumber \\
&&\left. -\  \frac{2}{\tilde m^2} e\left[ \left(\partial S\right)^2
- \frac{9}{4}\left(S^2 + \frac{1}{6}R\right)^2\right] +
\frac{1}{\mu}{\cal L}_{LCS}\right\}\;.
\end{eqnarray}
Note that the $S$ field is auxiliary in the limit that $\tilde
m^2\to\infty$, but with an equation that is not linear, in contrast
to the usual auxiliary fields of supergravity theories. Note also
that there is an $S^2R$ term, which means that elimination of $S$
could alter the `effective' curvature squared term in a vacuum with
non-zero $S$.
In this general model with both $L_K$ and
$L_{R^2}$ terms, there is a further very special case: that for
which \be\label{vsc} \tilde m^2 = 3 m^2\, . \ee This can be viewed
as the limit in which $\hat m^2\to\infty$ where the  mass parameter
$\hat m$ is defined by \be \frac{1}{\hat m^2} = \frac{1}{m^2} -
\frac{3}{\tilde m^2}\, . \ee The $S^2R$ and $S^4$ terms cancel in
the $\hat m^2\to\infty$ limit, and the curvature squared terms
become proportional to the square of the tracefree tensor
$R_{\mu\nu} - \frac{1}{3} g_{\mu\nu} R$.

In this section we will give the equations of motion, find some solutions and the amount of supersymmetry they preserve.
{}From (\ref{bosonicaction}) we find that the metric equation of
motion is
\begin{eqnarray}\label{metricEq}
0 &=& \left(- \frac{1}{2} MS + \sigma S^2\right) g_{\mu\nu} + \sigma G_{\mu\nu} + \frac{1}{\mu} C_{\mu\nu}  + \frac{1}{2m^2} K_{\mu\nu}
+ \frac{1}{2\tilde m^2} L_{\mu\nu} \nonumber \\
&&-\ \frac{2}{\tilde m^2} \left[\partial_\mu S \partial_\nu S - \frac{1}{2} g_{\mu\nu} \left(\partial S\right)^2\right] \nonumber \\
&&-\  \frac{1}{2\hat m^2} \left[G_{\mu\nu} S^2 -\frac{3}{2} g_{\mu\nu} S^4
-\left(D_\mu D_\nu - g_{\mu\nu} D^2\right)S^2\right] \, ,
\end{eqnarray}
where
\begin{eqnarray}
\sqrt{|g|}\, C_{\mu\nu} &=& \varepsilon_\mu{}^{\tau\rho} D_\tau S_{\rho\nu}\, , \qquad
S_{\mu\nu}= R_{\mu\nu}- \frac{1}{4}g_{\mu\nu}R\, ,\\
K_{\mu\nu} &=& 2D^2 R_{\mu\nu} - \frac{1}{2} D_\mu D_\nu R - \frac{1}{2} g_{\mu\nu} D^2R  - \frac{13}{8} g_{\mu\nu} R^2 \nonumber\\
&&+\  \frac{9}{2}R R_{\mu\nu}  -  8 R_\mu{}^\lambda R_{\lambda\nu} + 3 g_{\mu\nu} \left(R^{\rho\sigma}R_{\rho\sigma} \right)\;, \\
L_{\mu\nu} &=& - \frac{1}{2} D_\mu D_\nu R + \frac{1}{2} g_{\mu\nu}
D^2R - \frac{1}{8} g_{\mu\nu} R^2 + \frac{1}{2} RR_{\mu\nu}\;.
\end{eqnarray}
The tensor $C_{\mu\nu}$ is the Cotton tensor, which is a derivative
of the (3D) Schouten tensor $S_{\mu\nu}$; this term arises from
variation of the LCS term in the action.  The tensor $K_{\mu\nu}$ is
the tensor given in \cite{Bergshoeff:2009hq}; it arises from
variation of the $K$ term in the action. The tensor $L_{\mu\nu}$
arises from variation of the $R^2$ term in the action. The trace of
the metric equation can be written as
\begin{eqnarray}
&& \left(M-4\sigma S\right) S + 2\sigma\left(S^2 + \frac{1}{6} R\right) - \frac{1}{3m^2} \left(K+\frac{1}{2} S^2R
+ \frac{9}{2} S^4\right) \nonumber \\
&&\ + \frac{9}{2\tilde m^2} \left(S^2+ \frac{1}{6} R\right)
\left(S^2- \frac{1}{18} R\right) = \frac{2}{3\hat m^2} D^2 S^2 +
\frac{1}{3\tilde m^2}\left[ 2\left(\partial S\right)^2 + D^2
R\right]\;.
\end{eqnarray}
The $S$ equation of motion is
\be\label{SEq}
\left(M-4\sigma S\right) -
\frac{6}{\hat m^2} S\left(S^2+\frac{1}{6}R\right) = -\frac{4}{\tilde
m^2}D^2 S\, . \ee

\subsection{Maximally (super)symmetric vacua}\label{maxsusy}

We will now consider in detail the possibilities for maximally
symmetric, but not necessarily supersymmetric, vacua, for which $S$
is constant and \be G_{\mu\nu} = -\Lambda g_{\mu\nu} \ee for
cosmological constant $\Lambda$, which has dimensions of mass
squared.  For such solutions the condition (\ref{susycon2}) for
supersymmetry reduces to \be \Lambda + S^2 =0\, . \ee This was
derived as a necessary condition for supersymmetry but it is also
sufficient within the class of maximally symmetric vacua. Naturally,
it implies that $\Lambda\le0$ so that only Minkowski and adS vacua
can be supersymmetric.

The  $S$ equation of motion for  maximally symmetric solutions, with constant $S$, reduces to
\be\label{Seq}
\left(M- 4\sigma S\right) -\frac{6}{\hat m^2 } S\left(S^2+ \Lambda\right)=0\, .
\ee
Only the trace of the metric equation is needed, and this is
 \be\label{tracemetriceq}
 S\left(M- 4\sigma S\right) + \left(\Lambda+ S^2\right)\left[2\sigma + \frac{1}{2\hat m^2}
 \left(\Lambda-3S^2\right)\right] =0\, .
\ee
Note that both these equations simplify dramatically in the limit that $\hat m^2\to\infty$. In this special case
there is a unique vacuum for given $M$, with $S= M/(4\sigma)$  and $\Lambda= -M^2/16$. This vacuum is
Minkowski for $M=0$ and adS for  $M\ne0$, and supersymmetric in either case.  In the Minkowski vacuum the
linearized theory is non-unitary.

In a next step let us assume that $\hat m^2$ is finite, which
amounts to finding solutions for the generic curvature squared
theory. We observe that the equations (\ref{Seq}) and
(\ref{tracemetriceq}) imply that \be \left(M- 4\sigma S\right)
\left(9S^2 + 4\sigma \hat m^2 + \Lambda\right) =0\, . \ee This leads
to two branches of vacua. One comes from setting $M=4\sigma S$. In
this case $\Lambda +S^2=0$, so we have a supersymmetric vacuum when
$\Lambda<0$.  In a plot of $\Lambda$ against $M^2/16$, the vacua of
this branch  lie on a half-line, in the $\Lambda<0$ sector,  that
starts at the origin.

The other branch of vacua arises from solutions of  $9S^2 =-
\left(\Lambda + 4\sigma \hat m^2 \right)$. Substituting for $S$ in
(\ref{Seq}) we learn that \be \left(\Lambda+ 4\sigma \hat m^2\right)
\left(\Lambda + \frac{1}{4}\sigma \hat m^2\right)^2 +
\left(\frac{9\hat m^2 M}{16}\right)^2 =0\, , \ee which is a cubic
equation for $\Lambda$. Let us consider in turn the two possible
signs for $\sigma$:

\begin{itemize}
\item  $\sigma<0$. There is no solution  for $\Lambda$ unless
\be \Lambda < 4 \hat m^2\, . \ee If we plot $\Lambda$ against
$M^2/16$, we see that the cubic curve that gives the vacua on
this branch just touches the $M=0$ axis at $\Lambda =
\frac{1}{4}\hat m^2$. This means that  $M=0$ allows two dS vacua
(in addition to the supersymmetric Minkowski vacuum of the other
branch); one has $\Lambda= \frac{1}{4}\hat m^2$ and the other
has $\Lambda= 4\hat m^2$. These are connected in  the sense that
they lie on the same curve in the $M^2\ge0$ region of the
$(\Lambda, M^2/16)$ plane. This cubic curve also cuts the
$\Lambda=0$ axis, so there is a non-supersymmetric Minowski
vacuum (with non-zero $S$)  in addition to the supersymmetric
Minkowski vacuum on the other branch. The cubic curve intersects
the line $\Lambda+ M^2/16=0$ in two points given by
$M^2=8\hat{m}^2$ and $M^2=8(3\sqrt{3}-5)\hat{m}^2$, so these two
adS vacua are supersymmetric.

\item  $\sigma>0$. For $M=0$ there is a non-supersymmetric adS vacuum with $\Lambda=-\frac{1}{4}\hat m^2$.
This is `isolated' because the part of the cubic curve with
$M^2<0$ is unphysical. All other solutions on this branch are
such that \be \Lambda \le -4 \hat m^2\, . \ee The limiting
$\Lambda=-4\hat m^2$ adS vacuum occurs for $M^2=0$. All these
adS vacua  are non-supersymmetric, with one exception,
corresponding to the point in the $(\Lambda, M^2/16)$ plane at
which the cubic equation cuts the line of supersymmetric adS
vacua.

\end{itemize}

These possibilities are displayed in Figs.~1 and 2.
\begin{figure}
 \begin{center}
 \includegraphics[scale=0.9]{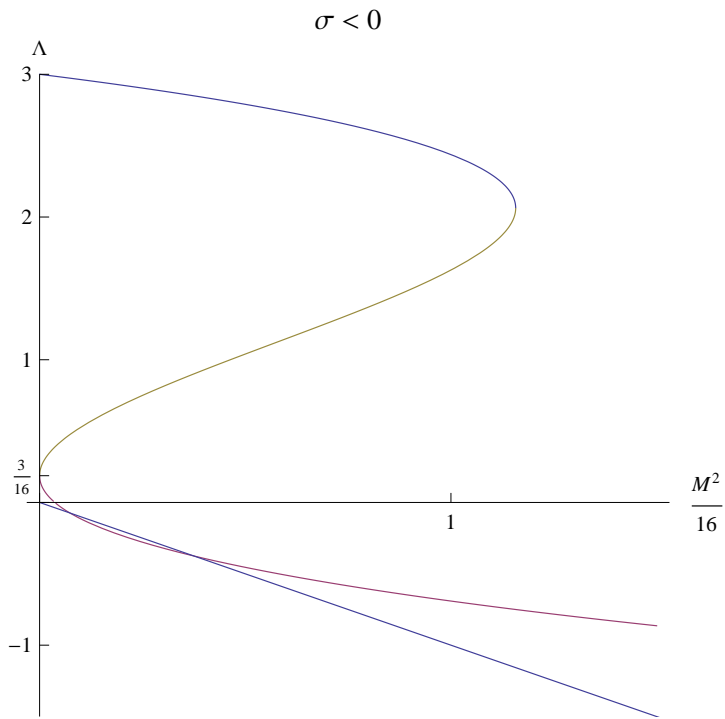}
 \label{graph1}
 \caption{Maximally-symmetric vacua for $\sigma=-1$ and $\hat{m}^2=\tfrac{3}{4}$, the straight
 line representing supersymmetric vacua.}
 \end{center}
\end{figure}

\begin{figure}
 \begin{center}
  \includegraphics[scale=0.9]{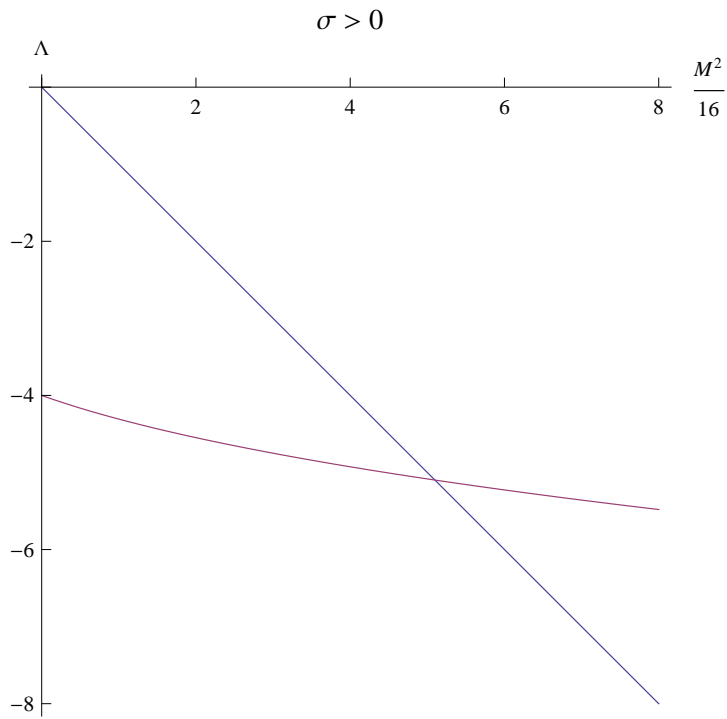}
  \caption{Maximally-symmetric vacua for $\sigma=1$ and $\hat{m}^2=1$, the straight line representing the
  supersymmetric vacua; this is the same straight line as in Fig.~1, despite appearances,  because of the different scales for the $\Lambda$ axes.  There is also an isolated adS vacuum at $\Lambda=-\ft14$, $M=0$, which is not indicated.}
 \end{center}
\end{figure}

Let us finally note that there is a neat geometrical interpretation
for the existence of supersymmetric adS solutions with
$\Lambda=-M^2/16$. Remarkably, this is precisely the value one gets
for pure Einstein-Hilbert plus cosmological constant. In other
words, the higher-derivative contributions to the field equations
drop out for maximally supersymmetric solutions. This can be
directly understood by noting that the connection $\Omega^{+}$ gives
rise to so-called `parallelizing torsion' for maximally
supersymmetric configurations. To be precise, from (\ref{RiemS}) we
infer that the curvature with respect to the torsionful connection
vanishes when evaluated for supersymmetric adS solutions,
 \bea\label{parallel}
  R_{\mu\nu}{}^{ab}(\Omega^{+})|_{S^2=-\Lambda} \ = \ 0 \;.
 \eea
Since the supersymmetric curvature-square actions have been computed
as the squares of $R(\Omega^{+})$, it follows directly from
(\ref{parallel}) that their contribution to the field equations
obtained by varying this action vanishes for supersymmetric
configurations. Explicitly, we have for the $K$ invariant the
factorization
 \bea
  L_{K} \ = \  \left(G^{\mu\nu}-S^2g^{\mu\nu}\right)
  \big[\left(G_{\mu\nu}-S^2g_{\mu\nu}\right)+\ft14(R+6S^2)g_{\mu\nu}\big]\;.
 \eea
Let us stress that since the first factor vanishes only for
maximally symmetric geometries, the variation of $L_{K}$ will not
vanish for general supersymmetric configurations. This is in
contrast to $L_{R^2}$ whose variation vanishes for all
supersymmetric solutions with constant $S$ by virtue of
(\ref{gensusycon}).

\subsection{pp-wave solutions}

We now aim to find a supersymmetric pp-wave metric (\ref{ppmetric})
that solves the metric equation (\ref{metricEq}) and $S$-equation
(\ref{SEq}). It is straightforward to verify that the $vv$ and $vx$
components of the metric equation  are automatically satisfied.
Thus, we have to consider the $uu$, $uv$, $ux$ and $xx$ components
of the metric equation, and the $S$-equation. The latter, upon the
use of (\ref{Sresult}) and (\ref{s3}), takes the form
\be \frac{1}{{\tilde m}^2}\,\partial_x^2 S  - \left(
\frac{1}{m^2}-\frac{1}{ {\tilde m}^2}\right) S\partial_x S - \sigma
S  + \frac14 M =0\ . \label{seom} \ee
Setting $S=-1/\ell$, we see from (\ref{seom}) that
$\ell=-4\sigma/M$. The non-vanishing components of the Ricci tensor
$R_{ab}$ and Cotton tensor $C_{ab}$ are
\bea
R_{+-} &=&R_{22}\ = \ -2\ , \\
R_{--} &=& -\frac{1}{2r^2} \left(r^2\partial_r^2
-r\partial_r\right)h\ ,
\label{e2} \\
C_{--} &=& (r\partial_r +1)R_{--}\ , \eea
where we set, from now on, $\ell=1$. Turning to the metric equations,
using these results and \eq{scc}, we find that they are all
trivially satisfied except the $uu$ component which takes the form
\bea\label{einstein} \left[
\frac{1}{m^2}r^2\partial_{r}^2+\left(\frac{3}{m^2}+\frac{1}{\mu}\right)r\partial_{r}
+\left(\sigma+\frac{1}{\mu}\right)\right] R_{--} \ = \ 0\ . \eea
To solve this equation, we substitute $h=r^n$. The resulting
characteristic polynomial is
\be n(n-2) \left( \frac1{m^2} n(n-2) + \frac1{\mu} (n-1)
+\sigma\right)=0\ . \label{ce} \ee
Thus we find the solutions
\bea h_\pm (u,r) \ = \ r^{n_\pm} f_{1}(u)+r^2f_{2}(u)+f_{3}(u)\ ,
\label{pp} \eea
where $f_{1,2,3}$ are arbitrary functions of $u$ only and
%
%\bea \label{ppn} n_\pm  \ = \ 1-\frac{m^2\ell}{2\mu}\pm
%\sqrt{1+\frac{m^4\ell^2}{4\mu^2}-\sigma m^2\ell^2}\ , \eea
%
%
\bea \label{ppn} n_\pm  \ = \ 1-\frac{m^2}{2\mu}\pm
\sqrt{1+\frac{m^4}{4\mu^2}-\sigma m^2}\ . \eea
The functions $f_2$ and $f_3$ can be removed by local coordinate
transformations (see, for example, \cite{Gibbons:2008vi}).
Therefore, we shall take $n_{\pm} \ne 0,2$ and write the general
solution as
\be h(u,r) = h_+ (u) r^{n_+} + h_-(u) r^{n_-}\ , \ee where
$h_{\pm}(u)$ are arbitrary functions of $u$, and the exponents
$n_{\pm}$ are as given in \eq{ppn}.

Next, we observe that in the bosonic NMG, the characteristic
equation obtained in \cite{AyonBeato:2009yq} has an additional
factor of $1/(2m^2)$ in the parenthesis multiplying $n(n-2)$ in
\eq{ce}. In the massive supergravity model we are considering,
however, there is an additional contribution coming from the term
proportional to $G_{\mu\nu} S^2$ in \eq{metricEq}. As a consequence,
we obtain the characteristic equation \eq{ce}, and the roots
\eq{ppn} differ from those in \cite{AyonBeato:2009yq} in that the
first term under the square root is $1$ instead of $\ft12$. This
difference has interesting consequences, as we shall see below.

To begin with, let us consider the roots of \eq{ppn} and examine the
parameter values for which degeneracies arise. In such cases, as is
well known, additional logarithmic solutions appear. The doubly
degenerate roots $n_+=n_-$ arise for
\be m^2_\pm = 2\mu^2 \left( \sigma \pm \sqrt{\sigma^2
-\frac1{\mu^2}}\right)\ , \ee where, again, we have suppressed the
adS radius, which can easily be re-introduced by dimensional
analysis,  for notational simplicity. In this case, the following
additional solutions arise
\be h(r,u) = r^{k_\pm} \left[h_1(u) \log r + h_2(u) \right]\ , \ee
where  $k_\pm  =  1-(m^2_\pm/ 2\mu) $ takes the form
\be k_\pm = 1-\mu\sigma \mp \sqrt{\mu^2\sigma^2-1}\ , \ee
and $h_1(u), h_2(u)$ are arbitrary functions of $u$.

Considering the root $n=0$ of \eq{ce}, it becomes triply degenerate
for $\mu\sigma=+1$, and the root $n=2$ becomes triply degenerate for
$\mu\sigma=-1$, since
\be
 k_{\pm} =
\begin{cases} 0 & \text{if $\mu\sigma=+1$ ,}
\\
2 &\text{if $\mu\sigma=-1$\ .}
\end{cases}
\ee
This means that the solutions becomes $adS_3$, and that the
following additional solutions arise:
\bea && \mu\sigma=+1:\qquad h(r,u)= \log r \left[ h_1(u) \log r +
h_2(u)\right]\ , \w2 && \mu\sigma=-1:\qquad h(r,u)= r^2 \log r
\left[ h_1(u) \log r + h_2(u)\right]\ . \eea
This is remarkable because $\mu\sigma=\pm 1$ are precisely the
critical points which arise in the chiral gravity limit of TMG
\cite{Li:2008dq} in which, apart from the logarithmic modes that do
not obey the standard Brown-Henneaux boundary conditions
\cite{Grumiller:2008qz},  the usual graviton mode ceases to
propagate in the bulk. In the case of ordinary bosonic NMG, on the
other hand, it can be shown that critical points arise for those
values at which the central charges of bosonic NMG vanish
\cite{AyonBeato:2009yq}. We shall comment further on various aspects
of our critical points $\mu\sigma=\pm 1$ in the conclusions.

%%%%%%%%%%%%%%%%%%%%%%%%%%%%%%%%%%%%%%%%%%%%%%%%%%%%%%%%%

\section{Linearization about a supersymmetric Minkowski vacuum}
\setcounter{equation}{0}

We now wish to investigate the propagating degrees of freedom and
their multiplet structure around a supersymmetric Minkowski vacuum,
which requires a linearisation about this background. Linearized 3D
${\cal N}=1$ supergravity theories are constructed from the
symmetric tensor $h_{\mu\nu}$, the anticommuting vector spinor
$\psi_\mu$ (which is a Majorana spinor) and the `auxiliary' scalar
$S$ (which may actually propagate, depending on the details of the
action). We insist on gauge invariance with respect to the following
linear gauge transformations \be h_{\mu\nu} \to h_{\mu\nu} +
\partial_{(\mu} v_{\nu)}\, , \qquad \psi_\mu \to \psi_\mu +
\partial_\mu \varsigma\, , \ee where $v$ is an arbitrary vector
field and $\varsigma$ an arbitrary Majorana spinor field. It is
convenient to define \be h_\mu = \eta^{\nu\rho} \partial_\rho
h_{\mu\nu}\, , \qquad h = \eta ^{\mu\nu} h_{\mu\nu} \, , \ee and to
introduce the following gauge invariant `field strengths' \be
R_{\mu\nu}^{(lin)} = - \frac{1}{2}\left[ \square h_{\mu\nu} -2
\partial_{(\mu}h_{\nu)} + \partial_\mu\partial_\nu h\right] \, ,
\qquad {\cal R}_{(lin)}^\mu = \varepsilon^{\mu\nu\rho} \partial_\nu
\psi_\rho\, . \ee The first of these is the linearized Ricci tensor
and the second is the Rarita-Schwinger field strength. The
linearized Einstein tensor is \be G_{\mu\nu}^{(lin)} =
R^{(lin)}_{\mu\nu} - \frac{1}{2} \eta_{\mu\nu} R^{(lin)}\, , \qquad
R^{(lin)} = \eta^{\mu\nu}R_{\mu\nu}^{(lin)}\, .\ee Also useful is
the linearized Cotton tensor \be C^{(lin)}_{\mu\nu} =
\varepsilon_\mu{}^{\tau\rho} \partial_\tau S^{(lin)}_{\rho\nu} \, ,
\qquad S^{(lin)}_{\mu\nu} = R_{\mu\nu}^{(lin)} - \frac{1}{4}
\eta_{\mu\nu} R^{(lin)}\, , \ee and its fermionic counterpart, the
`Cottino tensor', \be {\cal C}_{(lin)}^\mu = \gamma^\nu\partial_\nu
{\cal R}^\mu _{(lin)} + \varepsilon^{\mu\nu\rho} \partial_\nu {\cal
R}_\rho \, . \ee Note the identities \be \gamma_\mu {\cal
C}_{(lin)}^\mu \equiv 0\, , \qquad \partial_\mu {\cal C}^{\mu\nu}
\equiv 0 \, . \ee

The linearized off-shell supersymmetry transformations may now be written as
\begin{eqnarray}\label{susytrans}
\delta_\epsilon h_{\mu\nu} &=& \bar\epsilon\gamma_{(\mu}\psi_{\nu)} \, , \qquad
\delta_\epsilon S = \frac{1}{4} \bar\epsilon\gamma_\mu{\cal R}^\mu_{(lin)}\, ,
\nonumber \\
\delta_\epsilon \psi_\mu &=& \left[-\frac{1}{4} \varepsilon^{\rho\sigma\nu}\gamma_\nu\partial_\rho h_{\mu\sigma}
+ \frac{1}{2} S \gamma_\mu\right] \epsilon \, .
\end{eqnarray}
The following four quadratic Lagrangians yield actions that are both gauge invariant and  supersymmetric, up to surface terms:
\begin{eqnarray}
L_{EH}^{(2)} &=& -\frac{1}{2} h^{\mu\nu}G_{\mu\nu}^{(lin)}
-2S^2 -\bar\psi_\mu  {\cal R}^\mu_{(lin)}  \nonumber \\
L_{top}^{(2)} &=& \frac{1}{2} h_{\mu\nu} C^{\mu\nu} _{(lin)}  + \frac{1}{2} \bar\psi_\mu {\cal C}_{(lin)}^\mu \nonumber \\
 L_{K}^{(2)}  &=&  - \frac{1}{2} \varepsilon^{\mu\tau\rho}  h_{\mu}{}^\nu \partial_\tau C_{\rho\nu}^{(lin)}
 - \frac{1}{2} \bar\psi_\mu \left(\gamma^\nu\partial_\nu\right) {\cal C}_{(lin)}^\mu \nonumber \\
 L_{R^2}^{(2)} &=& R_{(lin)}^2 + 16 S\square S - 4 \left(\bar {\cal R}_{(lin)} \cdot \gamma\right)\left(\gamma \cdot\partial\right)\left(\gamma\cdot {\cal R}_{(lin)}\right) \, .
\end{eqnarray}
One can show that \be R^{\mu\nu}_{(lin)}R_{\mu\nu}^{(lin)} -
\frac{3}{8} R_{(lin)}^2 =  - \frac{1}{2} \varepsilon^{\mu\tau\rho}
h_{\mu}{}^\nu \partial_\tau C_{\rho\nu}^{(lin)}  + {\rm total \
derivative}\, , \ee so the Lagrangian $L_{K}^{(2)}$ is indeed the
quadratic approximation to the supersymmetrization of the Lagrangian
$L_K$.  Similarly, the Lagrangian $L_{top}^{(2)}$ is the quadratic
approximation to the LCS term since its variation yields the
linearized Cotton tensor.

In the following we shall  consider the general linear combination
of these four Lagrangians, which are parametrized by a dimensionless
constant $\sigma$ and three mass parameters $(\mu,m,\tilde m)$:
\be\label{lagcombo} L^{(2)} = \sigma L^{(2)}_{EH} + \frac{1}{\mu}
L^{(2)}_{top} + \frac{1}{m^2} L^{(2)}_{K} + \frac{1}{8\tilde m^2}
L^{(2)}_{R^2} \, . \ee On setting $\sigma=1$ and taking all mass
parameters to infinity,  one gets the linearization of the standard
${\cal N}=1$\ 3D supergravity, which has no propagating modes.
Allowing finite $\mu$ leads to a unitary theory if $\sigma<0$ and
one may then choose $\sigma=-1$ without loss of generality; this is
the linearization of topologically massive supergravity, which
propagates modes of helicities  $\pm(2,3/2)$, the sign depending on
the sign of $\mu$. Of principal interest here will be the models for
which either $m^2$ or $\tilde m^2$ is finite; as we shall see,
unitarity requires that we take either $m^2$ or $\tilde m^2$ to
infinity, but this is merely a necessary condition for unitarity,
not a sufficient one. Our aim here is to determine all possible
unitary theories within the class of models considered.

\subsection{Canonical decomposition}

There are three  gauge-invariant components of the metric, which we may write, following Deser \cite{Deser:2009hb} but in terms of slightly different variables $(N,\xi,\varphi)$ as
\be
h_{ij} = - \varepsilon^{ik} \varepsilon^{jl} \frac{\partial_k\partial_l}{\nabla^2} \varphi\, , \qquad
h_{0i} = -\varepsilon^{ij} \frac{1}{\nabla^2}\partial_j  \xi \,  , \qquad h_{00} = \frac{1}{\nabla^2} \left(N + \square\varphi\right)\, .
\ee
Observe that this decomposition implies the gauge choice
\be\label{gaugechoice1}
\partial_i h_{ij} \equiv 0 \, , \qquad \partial_i h_{0i} \equiv 0\, .
\ee
We may make a similar decomposition of the anticommuting vector spinor $\psi_\mu$ in terms of  anticommuting spinors $(\eta,\chi)$ by writing
\be
\psi_i = \gamma_i \chi\, , \qquad
\psi_0 = \gamma_0 \left(\frac{1}{\nabla^2}\gamma^i\partial_i \eta + \chi\right) \, .
\ee
This implies the gauge choice
\be\label{gaugechoice2}
\gamma^i \psi_i = 2 \frac{\gamma^i\partial_i}{\nabla^2} \partial_j \psi_j \, ,
\ee
which is non-standard but simplifies the subsequent analysis.

In terms of the variables $(N,\xi,\varphi)$, the components of the linearized Einstein tensor are
\begin{eqnarray}
G^{lin}_{00} &=& \frac{1}{2} \nabla^2\varphi \, , \qquad
G^{lin}_{0i} = \frac{1}{2}\left( \partial_i\dot\varphi + \epsilon^{ij}\partial_j\xi \right)\;,  \\
G^{lin}_{ij} &=& -\frac{1}{2} \left( \delta_{ij} \square \varphi - \partial_i\partial_j \varphi\right) - \frac{1}{2}\left(\delta_{ij} - \frac{\partial_i\partial_j}{\nabla^2} \right)N + \frac{1}{2} \left(\varepsilon^{ik} \frac{\partial_k\partial_j}{\nabla^2} +
\varepsilon^{jk} \frac{\partial_k\partial_i}{\nabla^2} \right)\dot\xi\, ,  \nonumber
\end{eqnarray}
and hence
\be
R^{lin} = N + 2\square\varphi\, .
\ee
The components of the linearized Cotton tensor are
\begin{eqnarray}
C_{00}^{(lin)} &=& \frac{1}{2} \nabla^2 \xi \, , \qquad
C_{0i}^{(lin)} = \frac{1}{2} \partial_i \dot\xi - \frac{1}{4} \varepsilon^{ij} \partial_j N\;, \\
C_{ij}^{(lin)} &=& \frac{1}{2} \left(\delta_{ij}\square + \partial_i\partial_j\right)\xi +
\frac{\partial_i\partial_j}{\nabla^2}\ddot\xi - \frac{1}{4} \left( \varepsilon^{ik}\frac{\partial_k\partial_j}{\nabla^2} +
\varepsilon^{jk}\frac{\partial_k\partial_i}{\nabla^2} \right)\dot N\, . \nonumber
\end{eqnarray}

In terms of the anticommuting spinor variables $(\eta,\chi)$, the components of the Rarita-Schwinger field strength are
\be
{\cal R}_{(lin)}^0 = \gamma^0 \gamma^i\partial_i \chi\, , \qquad
{\cal R}^i_{(lin)}  = \gamma_0 \left[ \varepsilon^{ij} \partial_j \left( \chi + \frac{\gamma^k\partial_k }{\nabla^2} \eta\right)
+ \gamma^i \dot\chi\right] \, ,
\ee
and hence
\be
\gamma_\mu {\cal R}^\mu_{(lin)} =  \gamma_0\eta + 2 \gamma^\mu\partial_\mu \chi\, .
\ee
The components of the fermionic counterpart of the Cotton tensor are
\be
{\cal C}^0 = \gamma^0 \gamma^i\partial_i \eta \, , \qquad {\cal C}^i = \varepsilon^{ij} \partial_j \left( \gamma^0 \eta + \frac{\gamma^k\partial_k}{\nabla^2} \dot \eta\right) -
\gamma^0 \partial_i \left[ \frac{\gamma^k\partial_k}{\nabla^2} \dot \eta\right] \, .
\ee

Using these results, one  finds that
\begin{eqnarray} \label{susylags}
L^{(2)}_{EH} &=& -\frac{1}{2} \left(\varphi N + \varphi \square \varphi - \xi^2\right) -2 S^2 + 2 \bar\chi \left(\gamma^\mu \partial_\mu\right) \chi +2 \bar\chi\eta\nonumber\;, \\
L^{(2)}_{top} &=&   \frac{1}{2} \xi N - \frac{1}{2}\bar\eta \eta \;, \\
L^{(2)}_{K} &=& \frac{1}{8} N^2 +  \frac{1}{2} \xi \square \xi  -\frac{1}{2}\bar\eta\left(\gamma^\mu \partial_\mu\right) \eta \nonumber\;, \\
L^{(2)}_{R^2} &=&  \left(N+ 2\square\varphi\right)^2  + 16  S\square S - 4 \bar\eta \left(\gamma^\mu\partial_\mu\right) \eta -16\bar\eta \, \square \chi - 16 \square\bar\chi \left(\gamma^\mu\partial_\mu\right) \chi\, . \nonumber
\end{eqnarray}
Notice that both $L^{(2)}_{top}$ and $L^{(2)}_{K}$ are independent
of both $\varphi$ and $\chi$. For $L^{(2)}_{top}$  this is a
consequence of its superconformal invariance. For $L^{2}_{K}$ it is
a consequence of an `accidental'  linearized superconformal
invariance that is not a feature of the full action.  The
combination of these Lagrangians corresponding to (\ref{lagcombo})
can be written as \be L^{(2)} = L^{(2)}_{(bos)} + L^{(2)}_{(ferm)}\,
, \ee where
\begin{eqnarray}\label{genbos}
L^{(2)}_{(bos)}   &=&  -\frac{\sigma}{2} \left(\varphi N + \varphi \square \varphi - \xi^2\right) +  \frac{1}{2\mu} \xi N
+ \frac{1}{8m^2} N^2 + \frac{1}{2m^2} \xi \square \xi  \nonumber \\
&& + \ \frac{1}{8\tilde m^2} \left(N+ 2\square\varphi\right)^2 +
\frac{2}{\tilde m^2} S\left(\square - \sigma\tilde m^2\right) S\, ,
\end{eqnarray}
and
\begin{eqnarray}\label{genferm}
L^{(2)}_{(ferm)}  &=&  2\sigma \left[ \bar\chi \gamma^\mu\partial_\mu \chi  + \bar\chi \eta \right] - \frac{1}{2\mu} \bar\eta\eta
- \frac{1}{2m^2} \bar\eta \gamma^\mu\partial_\mu \eta \nonumber \\
&& + \ \frac{1}{\tilde m^2} \left[- 4 \bar\eta
\left(\gamma^\mu\partial_\mu\right) \eta -16\bar\eta \, \square \chi
- 16\square\bar\chi \left(\gamma^\mu\partial_\mu\right)
\chi\right]\;.
\end{eqnarray}

A notable feature of the above Lagrangians is that they can be interpreted as Lorentz invariant Lagrangians in their own right, despite the initial time-space split that was used to arrive at them. In this context, we would interpret the bosonic fields as Lorentz scalars and the fermionic fields as Lorentz spinors. However, the stress tensor of this scalar-spinor theory is not the same as that of the `original' theory, and hence the integral for angular momentum is quite different to that of the original theory, so one cannot read off the spins of the propagated modes in the original theory in any obvious way. However, the formalism is well-suited to the task of determining all possible unitary theories. Once we have these theories, other methods must be used to determine the helicity content (in the case of massive modes, because helicity is not defined for massless particles in 3D).

\subsubsection{Check of supersymmetry}

To determine the supersymmetry transformations of the  variables
$(N,\xi,\varphi)$ and $(\eta,\chi)$, we must consider the combined
transformations \be \delta h_{\mu\nu} = \delta_\epsilon h_{\mu\nu} +
\partial_{(\mu}^{}v^{(comp)}_{\nu)}\, , \qquad \delta\psi_\mu =
\delta_\epsilon \psi_\mu + \partial_\mu \varsigma^{(comp)}\, , \ee
where the $\delta_\epsilon$ variations are those of
(\ref{susytrans}) and the parameters  of the (compensating) gauge
transformations must be chosen such that the combined
transformations  preserve the gauge choices (\ref{gaugechoice1}) and
(\ref{gaugechoice2}). This requirement implies that \be v^{(comp)}_0
= \bar\epsilon\frac{1}{\nabla^2}\left( \gamma_0 \eta +
\dot\chi\right)\, , \qquad v_i^{(comp)} = - \bar\epsilon
\frac{1}{\nabla^2} \partial_i \chi\, , \ee and that \be
\varsigma^{(comp)} = -\frac{1}{4}\left[ \varphi  + \gamma^i
\partial_i \frac{1}{\nabla^2} \left(\gamma^0 \dot\varphi -
\xi\right)\right]\epsilon\, . \ee One then finds that
\be\label{transcan1} \delta N = - \bar\epsilon\gamma^\mu
\partial_\mu \eta\, , \qquad \delta\xi = -\frac{1}{2}\bar\epsilon
\eta\, , \qquad \delta\varphi = -\bar\epsilon\chi\, , \qquad \delta
S = \frac{1}{2} \bar\epsilon \gamma^\mu \partial_\mu \chi +
\frac{1}{4} \bar\epsilon \eta\, , \ee and that \be\label{transcan2}
\delta \chi = -\frac{1}{4} \gamma^\mu\epsilon \partial_\mu\varphi +
\frac{1}{4} \xi\epsilon + \frac{1}{2} S\epsilon\, , \qquad \delta
\eta = -\frac{1}{4} N \epsilon -\frac{1}{2} \gamma^\mu\epsilon
\partial_\mu \xi\, . \ee One may verify that all four Lagrangians
(\ref{susylags}) are invariant under these transformations.

\subsection{Unitarity}

We now use the above results to find all unitary theories within the
class of the theories parametrized by $(\sigma,\mu,m,\tilde m)$. We
shall do this separately for the bosonic part and the fermionic
bilinear part.

\subsubsection{Bosonic part}

The $N$ field is auxiliary in (\ref{genbos}) and can be eliminated
to yield the equivalent Lagrangian\footnote{There are special cases
for which $N$ occurs only linearly, in which case it is a Lagrange
multiplier for a constraint, but the solution of the constraint
turns out to yield models  that can also be obtained as limits of
the generic ones obtained by integrating out $N$.}
\begin{eqnarray}
L^{bos} &=&
 \frac{1}{2\left(m^2+\tilde m^2\right)} \left(\square\varphi\right)^2 + \frac{1}{2m^2} \xi \square\xi -
  \frac{m^2}{\left(m^2+\tilde m^2\right)\mu}\xi \square\varphi - \frac{\sigma}{2} \left(\frac{\tilde m^2 - m^2}{\tilde m^2 + m^2} \right) \varphi \square\varphi \nonumber \\
&& -\  \frac{m^2\tilde m^2}{2\left(m^2+\tilde m^2\right)} \left[ \sigma^2\varphi^2 - 2\frac{\sigma}{\mu} \varphi \xi +
\left(\frac{1}{\mu^2} - \frac{\left(m^2+\tilde m^2\right)\sigma}{m^2\tilde m^2}\right) \xi^2\right] \nonumber \\
&& +\  \frac{2}{\tilde m^2} S\left(\square - \sigma\tilde m^2\right) S\, .
\end{eqnarray}
There are ghosts unless the  $(\square\varphi)^2$ term is absent, which requires that $m^2+ \tilde m^2 \to \infty$.  We may take $\tilde m^2\to \infty$ keeping $m^2$ fixed, or {\it vice-versa}. We shall consider these two possibilities in turn

\begin{itemize}

\item $\tilde m^2\to \infty$. In this case it is convenient to set
\be
\xi = m\zeta\, ,
\ee
after which the Lagrangian becomes
\be
L^{bos}=  \frac{1}{2} \left[- \sigma \varphi \square \varphi + \zeta \square\zeta\right]
 - \frac{1}{2} m^2\left[\sigma^2 \varphi^2 \mp  2 \sigma \frac{m}{\mu} \varphi\zeta + \frac{\left(m^2 -\sigma \mu^2\right)}{\mu^2} \zeta^2\right] \nonumber\;. \\
\ee
This result generalizes that of \cite{Deser:2009hb} to allow for $\sigma\ne-1$ and $|\mu|\ne\infty$.  We see that
$\sigma\le0$ is necessary for unitarity.

Consider first the $\sigma<0$ case; we may then choose $\sigma=-1$
without loss of generality. In terms of the row 2-vector
$\Phi^T=(\varphi,\zeta)$, the Lagrangian takes the form \be L^{bos}
= \frac{1}{2} \Phi^T \square \Phi - \frac{1}{2} \Phi^T M^2 \Phi\, ,
\ee where $M^2$ is a mass matrix with eigenvalues $m^2_\pm$ such
that \be m_+ m_- = m^2\, , \qquad |m_+-m_-| =  \frac{m^2}{|\mu|}\, .
\ee We thus find agreement with \cite{Bergshoeff:2009hq}, although
it is not obvious from this analysis that both modes have spin 2.

When $\sigma=0$ we get the Lagrangian \be L^{bos} =
\frac{1}{2}\left[ \zeta \square \zeta -
\left(\frac{m^2}{\mu}\right)^2 \zeta^2\right] - \frac{1}{2} m^2
\varphi^2\;. \ee The variable $\varphi$ is now auxiliary so we
have a single mode with mass $m^2/\mu$; it will be shown that
this mode has spin 2, so the model is, at least at the
linearized level, a `new topologically massive gravity' (NTMG).

\item $m^2\to\infty$. In this case we have
\begin{eqnarray}\label{SMG}
L^{bos} &=& -\frac{1}{\mu} \xi \square \varphi + \frac{\sigma}{2} \varphi \square \varphi
- \frac{1}{2}\tilde m^2 \left[ \sigma^2 \varphi^2 - 2\frac{\sigma}{\mu} \, \varphi \xi + \left(\frac{1}{\mu^2} - \frac{\sigma}{\tilde m^2} \right)\xi^2\right] \nonumber \\
&&+ \  \frac{2}{\tilde m^2} S\left(\square - \sigma\tilde m^2\right) S\, .
\end{eqnarray}
Given that $\sigma\ne0$, we may simplify  the Lagrangian by using  the new variables $(\varphi', \zeta')$ defined by
\be
\varphi = \varphi' \mp \zeta'/\sigma \, , \qquad \xi = -\mu\zeta'\, .
\ee
One then finds that
\be
L^{bos} =
 \frac{\sigma}{2} \varphi' \left(\square - \sigma \tilde m^2\right) \varphi'
 - \frac{1}{2\sigma} \zeta'\left(\square - \sigma^2\mu^2\right)\zeta'
 + \frac{2}{\tilde m^2} S\left( \square - \sigma \tilde
 m^2\right)S\;.
 \ee
We see that either $\varphi'$ or $\zeta'$ is a ghost mode, but we can still get a unitary theory by taking the ghost mass to infinity.
Returning to (\ref{SMG}) and taking  $\mu^2\to\infty$ we get the Lagrangian
\be
L^{bos} =  \frac{\sigma}{2} \varphi \left(\square - \sigma \tilde m^2\right) \varphi +  \frac{2}{\tilde m^2} S\left(\square - \sigma\tilde m^2\right) S+ \frac{\sigma}{2\tilde m^2} \xi^2 \, .
\ee
The variable $\xi$ is now auxiliary and may be trivially eliminated, resulting in a theory that is unitary and tachyon-free for $\sigma>0$; we may choose $\sigma=1$ without loss of generality. This unitary `scalar massive gravity' (SMG) theory propagates  two  scalar modes of mass  $\tilde m$; one mode comes from the metric and the other comes from the `auxiliary' scalar $S$.

If $\sigma=0$ then (\ref{SMG}) becomes
\be
L^{bos} = -\frac{1}{\mu} \xi \square\varphi - \frac{\tilde m^2}{2\mu^2} \xi^2 + \frac{2}{\tilde m^2} S\square S\, .
\ee
We see that $\xi$ is auxiliary again, but its elimination now yields the non-unitary Lagrangian
\be
L^{bos} = \frac{1}{2\tilde m^2} \left(\square\varphi\right)^2 + \frac{2}{\tilde m^2} S\square S\, .
\ee

\end{itemize}

To summarize, there are essentially just three ways to get a unitary Lagrangian when either $m^2$ or $\tilde m^2$ is finite. These are
\begin{enumerate}

\item $\tilde m^2\to\infty$ and $\sigma=-1$. This yields GMG.

\item $\tilde m^2\to \infty$ and $\sigma=0$. This yields `New Topologically massive gravity'' (NTMG),
but this model may have problems at the interacting level. The
massless version is the `pure-K' model considered by Deser
\cite{Deser:2009hb}.

\item $m^2\to\infty$ {\it and}  $\mu^2\to\infty$, and $\sigma=1$. This is the bosonic sector of SMG; it is equivalent to 3D gravity coupled to a scalar field with a particular potential that linearizes to give a particle of mass $\tilde m$, plus an `auxiliary' scalar describing another
particle of mass $\tilde m$.

\end{enumerate}

\subsubsection{Fermionic part}

It is convenient to rewrite the $1/\tilde m^2$ contribution to  (\ref{genferm}) so that
\begin{eqnarray}\label{genfermalt}
L^{(2)}_{(ferm)} &=& -\frac{1}{2m^2} \bar\eta \gamma^\mu\partial_\mu \eta - \frac{1}{2\tilde m^2} \bar\beta \gamma^\mu\partial_\mu \beta
+2\sigma \bar\chi \gamma^\mu\partial_\mu \chi - \frac{1}{\tilde m^2} \bar\lambda \gamma^\mu\partial_\mu \chi \nonumber \\
&&+\ 2\sigma \bar\chi \eta - \frac{1}{2\mu} \bar\eta\eta  - \frac{1}{2\tilde m^2} \bar\lambda \left(\eta - \beta\right) \, .
\end{eqnarray}
This involves two new spinor variables $(\beta,\lambda)$ but  $\lambda$ is a Lagrange multiplier that imposes the constraint $\beta= \eta + 2\gamma^\mu\partial_\mu\chi$,  whereupon the Lagrangian reduces to the previous one of (\ref{genferm}).

The kinetic terms for $(\chi,\lambda)$ can be brought to diagonal form in new variables but the result is that there is a ghost unless
{\it either}  (i) $\tilde m^2\to\infty$ {\it or} (ii) $m^2\to\infty$ {\it and} $\mu^2\to\infty$.  We shall consider in turn these two possibilities.

\begin{itemize}

\item $\tilde m^2\to\infty$. The fermionic Lagrangian simplifies to
\be\label{GMGferm} L^{ferm} =  -\frac{1}{2m^2} \bar\eta
\gamma^\mu\partial_\mu \eta +2\sigma \bar\chi
\gamma^\mu\partial_\mu \chi +2\sigma \bar\chi \eta -
\frac{1}{2\mu} \bar\eta\eta\;.
\ee Unitarity requires $\sigma<0$
and we may choose $\sigma=-1$ without loss of generality. By
setting \be \eta = m\eta' \, , \qquad \chi = \frac{1}{2}\chi' \,
, \ee and introducing a row 2-vector $\Xi^T= (\eta',\chi')$, we
can put the Lagrangian in the form \be\label{lagform} L^{ferm} =
-\frac{1}{2} \bar\Xi \left(\gamma^\mu\partial_\mu -M\right)\Xi
\, , \ee where $M$ is a diagonalizable mass matrix such that \be
\det M^2 = m^4\, ,\qquad  {\rm tr} M^2 = \frac{m^2\left(m^2+
2\mu^2\right)}{\mu^2} \, . \ee This implies that $M^2$ has
eigenvalues $m^2_\pm$, the squared masses of GMG.  Supersymmetry
implies that the two propagated  modes have spin $3/2$, but this
fact is not obvious from this approach.

When $\sigma=0$ the Lagrangian (\ref{GMGferm}) simplifies to
\be
L^{ferm} =  -\frac{1}{2} \bar\eta' \left(\gamma^\mu\partial_\mu + \frac{m^2}{\mu}\right)\eta'  \, .
\ee
This is the fermionic part of NTMG. As expected, it propagates a single mode of mass $m^2/\mu$. Supersymmetry implies that this mode has spin $3/2$.

\item $m^2\to\infty$ and $\mu^2\to\infty$.  Taking the limit $m^2\to\infty$ does not immediately remove the ghost modes from (\ref{genferm}) but it removes the kinetic term for $\eta$. If we also remove the mass term by taking $|\mu|\to\infty$ then $\eta$ becomes a Lagrange multiplier for the constraint
\be \lambda = 4\tilde m^2 \sigma \chi\, . \ee Using this we
arrive at the Lagrangian \be L^{ferm} = -\frac{1}{2\tilde m^2}
\bar\beta \gamma^\mu\partial_\mu\beta - 2\sigma \bar\chi
\gamma^\mu\partial_\mu \chi +2\sigma \bar\beta\chi\;. \ee We now
see that unitarity also requires $\sigma\ge0$. When $\sigma>0$
we may choose $\sigma=1$ without loss of generality. By setting
\be \beta= \tilde m \beta' \, , \qquad  \chi = \frac{1}{2}\chi'
\, , \ee and again introducing a row 2-vector $\Xi^T=
(\eta',\chi')$, we can again put the Lagrangian in the form
(\ref{lagform}) but now with a mass matrix $M$ such that $M^2$
has both eigenvalues equal to $\tilde m^2$. This is to be
expected because in the supersymmetrization of SMG the
`auxiliary' scalar $S$ propagates with mass $\tilde m$, so we
need two spin $1/2$ modes of this mass.

When $\sigma=0$, we get the very simple Lagrangian \be L^{ferm}
= -\frac{1}{2} \bar\beta' \gamma^\mu\partial_\mu \beta'\;, \ee
which propagates a single massless mode. This is the
superpartner to the `Deser' mode of the `pure-K' theory.

\end{itemize}

To summarize, the fermionic Lagrangian provides exactly the modes
implied by supersymmetry given our earlier bosonic results.

%%%%%%%%%%%%%%%%%%%%%%%%%%%%%%%%%%%%%

\section{The three unitary theories}
\setcounter{equation}{0}

Our investigations so far can be summarized by saying that among the generic `higher-derivative'  supergravity theories there are three classes of unitary theories:
\begin{itemize}

\item GMSG or `General Massive Supergravity'. This is obtained by setting $\sigma=-1$ and $\tilde m^2=\infty$, so that
\be\label{GMGact} I_{GMSG} \ = \ \frac{1}{\kappa^2} \int \! d^3x
\left\{e \left[-  L_{EH} + \frac{1}{m^2} L_K \right] +
\frac{1}{\mu}{\cal L}_{LCS}\right\} + fermions\;. \ee This
includes the supersymmetric extensions of both  `New Massive
Gravity' (NMG) and `Topologically Massive Gravity' (TMG),
obtained as the limiting cases in which $\mu^2\to\infty$ or
$m^2\to\infty$, respectively.

\item NTMSG or `New Topologically Massive Supergravity'. This is obtained by setting $\sigma=0$ and $\tilde{m}^2=\infty$, and so
\be\label{NTMGact} I_{NTMSG} \ = \ \frac{1}{\kappa^2} \int \!
d^3x \left\{\frac{1}{m^2} e\, L_K+ \frac{1}{\mu}{\cal
L}_{LCS}\right\} + fermions\;. \ee The bosonic action might be
considered as a limit of GMG in which $\sigma\to0$ but  there
are various reasons for considering it separately. In contrast
to NMG and TMG, one cannot get to the theory with $\sigma=0$
just by taking  limits of particle masses.  Also, there is an
`accidental'  superconformal invariance of the linearized theory
when $\sigma=0$, and this  means that the quadratic
approximation leads to a linearized Minkowski space field theory
with a `missing'  field equation. Interpretation of the
linearized results is therefore not straightforward.
Nevertheless, we will show here that this linearized theory has
many features in common with  TMG, hence the name we choose for
it. In particular, it propagates a single spin 2 mode, and its
fermionic counterpart propagates a single spin  3/2 mode.

\item SMSG or `Scalar Massive Supergravity'. This is obtained by setting $\sigma=1$ and both $\mu=\infty$ and $m^2=\infty$, so that
\be\label{SMGact} I _{SMSG}\ = \ \frac{1}{\kappa^2} \int \! d^3x
\ e \left[L_{EH}  + \frac{1}{8\tilde m^2} L_{R^2}\right]  +
fermions \, . \ee In the context of the purely bosonic theory,
and ignoring the supergravity `auxiliary' field $S$, this is
known to be equivalent to a scalar field coupled to gravity with
a potential that gives the scalar field a mass $\tilde m$ in the
linearized limit (see \cite{Schmidt:2006jt} for a review).  This
model has never been supersymmetrized, to our knowledge.

\end{itemize}

We shall now consider in turn these three classes of unitary supergravity theories and determine the
 helicities of the different fields.

\subsection{General Massive Supergravity}

The quadratic approximation to the Lagrangian of the `general massive supergravity' model is
\be\label{GMSG}
L^{(2)}_{GMSG} = L^{(2)}_{(bos)} + L^{(2)}_{(ferm)}\, ,
\ee
where
\begin{eqnarray}
L^{(2)}_{(bos)} &=& \frac{1}{2} h^{\mu\nu}G_{\mu\nu}^{(lin)} + 2S^2 +  \frac{1}{2\mu} h_{\mu\nu} C^{\mu\nu} _{(lin)}
- \frac{1}{2m^2} \varepsilon^{\mu\tau\rho}  h_{\mu}{}^\nu \partial_\tau C_{\rho\nu}^{(lin)} \;,\nonumber \\
L^{(2)}_{(ferm)} &=& \bar\psi_\mu {\cal R}_{(lin)}^\mu + \frac{1}{2\mu} \bar\psi_\mu {\cal C}^\mu
- \frac{1}{2m^2} \bar\psi_\mu \left(\gamma^\nu\partial_\nu\right) {\cal C}_{(lin)}^\mu\;.
\end{eqnarray}
The field $S$ is genuinely auxiliary and may be trivially eliminated. It was observed in \cite{Bergshoeff:2009hq} that the metric perturbation field equation can be written as
\be\label{GMGeqs}
\left[{\cal O}\left( -m_-\right) {\cal O}\left(m_+\right) \right]_\mu{}^\rho G^{(lin)}_{\rho\nu} =0 \, , \qquad  R^{(lin)}=0\, ,
 \ee
where the masses $m_\pm$ are given by \be m^2 = m_+m_- \, , \qquad
\mu = \frac{m_+m_-}{\left(m_--m_+\right)}\, , \ee and  ${\cal O}$ is
the following operator, defined for arbitrary mass $m$:
\be\label{sqrtProca} \left[{\cal O}(m)\right]_\mu{}^\nu \equiv
\delta_\mu{}^\nu + \frac{1}{m} \varepsilon_\mu{}^{\tau\nu}
\partial_\tau\, . \ee
Because of the linearized Bianchi identity $\partial^\mu
G^{lin}_{\mu\nu}=0$, the equations (\ref{GMGeqs}) propagate two spin
2 modes, with masses $m_+$ for helicity $+2$ and mass $m_-$ for
helicity $-2$. Here we shall present a novel proof of this fact .

Consider first the special case with $m_+=m_-$; in this case we need to prove that
the equations (\ref{GMGeqs}) are equivalent to the 3D version of the standard
Fierz-Pauli (FP) equation \cite{FP}. Actually, Fierz and Pauli presented their results
in terms of one dynamical equation and two subsidiary conditions. For a 3D symmetric tensor field $\tilde h$,
these equations are
\be\label{FPthree}
\left(\square - m^2\right) \tilde h_{\mu\nu} =0 \, , \qquad
\eta^{\mu\nu}\tilde h_{\mu\nu}=0\;, \qquad \partial^\mu \tilde
h_{\mu\nu} =0 \, . \ee
We may solve the differential subsidiary
condition by writing \be \tilde h_{\mu\nu} =
G^{(lin)}_{\mu\nu}(h)\;, \ee
where $G^{(lin)}$ is the linearized
Einstein tensor for a new symmetric tensor field $h$. The remaining
subsidiary constraint and the dynamical equation are, when expressed
as equations for $h$, precisely those of (\ref{GMGeqs}) in the
special case that $m_+=m_-$.  This proves the equivalence of  linearized NMG to the
3D FP theory.  To obtain the analogous result for GMG, one must start from the parity-violating modification
of the 3D FP equation found by replacing the wave equation for
$\tilde h$ with the equation \be \left[{\cal O}(-m_-) {\cal
O}(m_+)\right]_\mu{}^\rho \tilde h_{\rho\nu} =0\, .
\ee

Given this result for the bosonic Lagrangian, supersymmetry implies
that the two modes of masses $m_\pm$ propagated by the  fermionic
Lagrangian must have either spin 3/2 or spin 5/2.  We shall now show
that these modes have spin 3/2. The $\psi_\mu$ field equation is
\be\label{genericRSvar} {\cal R}_{(lin)}^\mu  +  \frac{1}{2\mu} {\cal
C}^\mu_{(lin)} - \frac{1}{2m^2}
\left(\gamma^\tau\partial_\tau\right) {\cal C}^\mu_{(lin)} = 0\;.
\ee Observe that this equation implies that \be \gamma \cdot {\cal
R}_{(lin)} =0\, . \ee To go further it is convenient to consider
first the limiting case in which $m^2\to \infty$: in this case we
have the equation \be\label{TMGferm}
\left(\gamma^\tau\partial_\tau\right) {\cal R}^\mu_{(lin)} = -2\mu
{\cal R}^\mu_{(lin)} - \varepsilon^{\mu\nu\rho}\partial_\nu{\cal
R}_\rho^{(lin)}  \qquad (m^2=\infty)\;, \ee which can be written as
\be \left[\hat {\cal O}(\mu) {\cal R}\right]^\nu =0\, , \ee where
\be \hat {\cal O}(\mu)^\mu{}_\nu = \delta^\mu{}_\nu +  \frac{1}{2\mu}
\left[ \delta^\mu{}_\nu \left(\gamma^\tau\partial_\tau\right) -
\varepsilon_\nu{}^{\tau\mu} \partial_\tau\right]\, . \ee We know
from studies of super-TMG that this equation must propagate  a
single spin 3/2 mode of mass $\mu$ \cite{Deser:1982sw,Deser:1984py}.
Next, we observe that the generic field equation
(\ref{genericRSvar}) can be written in the form \be \left[\hat {\cal
O}(-m_-)\hat {\cal O}(m_+)\right]^\mu{}_\nu\,  {\cal R}^\nu_{(lin)} =0
\;. \ee There is a precise parallel with our analysis of the spin 2
equation of GMG, as expected from supersymmetry.  The helicity $+2$
propagated with mass $m_+$ is accompanied by a helicity
$+\frac{3}{2}$ mode of the same mass, and the same for the negative
helicity states but with mass $m_-$.

\subsection{New Topologically Massive Supergravity}

The quadratic approximation to the Lagrangian of the `new topologically massive supergravity' model is
\be\label{GMSG}
L^{(2)}_{NTMSG} = L^{(2)}_{(bos)} + L^{(2)}_{(ferm)}\, ,
\ee
where
\begin{eqnarray}
L^{(2)}_{(bos)} &=&  \frac{1}{2\mu} h_{\mu\nu} C^{\mu\nu} _{(lin)}
- \frac{1}{2m^2} \varepsilon^{\mu\tau\rho}  h_{\mu}{}^\nu \partial_\tau C_{\rho\nu}^{(lin)}\;, \nonumber \\
L^{(2)}_{(ferm)} &=&   \frac{1}{2\mu} \bar\psi_\mu {\cal
C}_{(lin)}^\mu - \frac{1}{2m^2} \bar\psi_\mu
\left(\gamma^\nu\partial_\nu\right) {\cal C}_{(lin)}^\mu \;.
\end{eqnarray}
As we have seen, this model propagates one bosonic mode and one fermionic mode, both of mass
\be
\tilde \mu = m^2/\mu\, .
\ee
We now show that these modes have spin 2 and spin 3/2 respectively.

The linearized field equation for $h$ can be written as
\be\label{NTMGeq} \left[{\cal O}(\tilde\mu)\right]_\mu{}^\rho
C^{lin}_{\rho\nu} =0 \, , \qquad \left[{\cal
O}(\tilde\mu)\right]_\mu{}^\nu = \delta_\mu{}^\nu +
\frac{1}{\tilde\mu} \varepsilon_\mu{}^{\tau\nu} \partial_\tau\, .
\ee The tensor operator ${\cal O}(\tilde\mu)$ is the `square-root'
of the `Proca' operator \cite{Townsend:1983xs}. Despite appearances,
the tensor ${\cal O}(M)C^{lin}$ is symmetric by virtue of the
tracelessness of $C^{(lin)}$ and the `Bianchi' identity
\be\label{idens}
\partial^\mu C^{lin}_{\mu\nu} \equiv 0\, .
\ee As a consequence of this identity, we have the further identity
\be -\tilde\mu^2 \left[{\cal O}(-\tilde\mu){\cal O}(\tilde\mu)
C^{lin}\right]_{\mu\nu}  \equiv \left(\square -
\tilde\mu^2\right)C_{\mu\nu}\, , \ee from which it follows that that
the field equation ${\cal O}(\tilde\mu)C^{lin}=0$ implies that
\be\label{PFC} \left(\square - \tilde\mu^2\right)C^{lin}_{\mu\nu}
=0\, . \ee The combination of this equation with  (\ref{idens}) is
equivalent to the FP equation for the symmetric tensor $C^{lin}$.
This is not the independent field, of course, but this does not
matter because the equation $C^{lin}=0$ implies that $h$ is pure
gauge. One may expand on this argument along the lines presented for
NMG in  \cite{Bergshoeff:2009hq}, but here we present an alternative
argument that extends the one used above for GMG. Starting with  the
FP equations in the form (\ref{FPthree}) for the symmetric tensor
field $\tilde h$, we may solve {\it both} of the subsidiary
conditions by writing \be \tilde h_{\mu\nu} =
C^{(lin)}_{\mu\nu}(h)\;, \ee where $ C^{(lin)}(h)$ is the Cotton
tensor for a new symmetric tensor field $h$. The remaining dynamical
equation for $\tilde h$ is, when expressed as an equation for $h$,
precisely (\ref{PFC}).

We now turn to the linearized equation for the vector spinor field:
\be\label{RSC} \left(\gamma^\nu\partial_\nu -  \tilde\mu\right) {\cal
C}^\mu =0\, . \ee This propagates spin 3/2 because the spin 1/2
components are absent as a consequence of the identities
$\partial\cdot {\cal C}\equiv 0$ and $\gamma \cdot {\cal C}\equiv
0$.

\subsection{Scalar Massive Supergravity}

The quadratic approximation to the Lagrangian of the `scalar massive supergravity' model is
\be\label{GMSG}
L^{(2)}_{SMSG} = L^{(2)}_{(bos)} + L^{(2)}_{(ferm)}\, ,
\ee
where
\begin{eqnarray}
L^{(2)}_{(bos)} &=& -\frac{1}{2} h^{\mu\nu}G_{\mu\nu}^{(lin)} -2S^2
+ \frac{1}{\tilde m^2} \left[\frac{1}{8}R_{(lin)}^2  + 2 S\square S\right]\;, \nonumber \\
L^{(2)}_{(ferm)} &=& -\bar\psi_\mu  {\cal R}^\mu_{(lin)}  - \frac{1}{2\tilde m^2} \left(\bar {\cal R}_{(lin)} \cdot \gamma\right)\left(\gamma \cdot\partial\right)\left(\gamma\cdot {\cal R}_{(lin)}\right) \, .
\end{eqnarray}
In this case the field $S$ is not actually auxiliary; it propagates
a spin zero mode of mass $\tilde m$. It is known that the one mode
of mass $\tilde m$ propagated by the metric part of the bosonic
Lagrangian also has spin zero, so supersymmetry implies that the
fermionic part must propagate two spin 1/2 modes of mass $\tilde m$.
To verify this, we rewrite the `fermionic' Lagrangian as \be
L^{(2)}_{(ferm)} = - \bar\psi_\mu {\cal R}^\mu_{(lin)} -
\frac{1}{2}\bar \rho \gamma^\tau\partial_\tau \rho +
\bar\lambda\left(\tilde m \rho - \gamma\cdot {\cal
R}_{(lin)}\right)\;, \ee where the new spinor field $\lambda$ is a
Lagrange multiplier field that constrains the other new spinor field
$\rho$ to equal $\gamma\cdot {\cal R}_{(lin)}/\tilde m$. The general
solution of the $\psi_\mu$ field equation is \be \psi_\mu =
\frac{1}{2} \gamma_\mu \lambda + \partial_\mu\epsilon\, . \ee Thus,
$\psi$ is determined in terms of $\lambda$ up to an irrelevant gauge
transformation. Using this result, the $\lambda$ equation becomes
\be \gamma^\tau\partial_\tau \lambda = \tilde m \rho\, , \ee while
the $\rho$ field equation is \be \gamma^\tau\partial_\tau \rho=
\tilde m \lambda\, . \ee It follows that \be
\left(\gamma^\tau\partial_\tau \pm \tilde m \right)\left(\lambda\pm
\rho\right) = 0\, . \ee
 which implies two spin 1/2 modes of mass $\tilde m$.

%%%%%%%%%%%%%%%%%%%%%%%
\section{${\cal N}>1$ massive supergravities}
\setcounter{equation}{0}

Our results for ${\cal N}=1$ 3D supergravities can be extended to ${\cal N}=2$. The linearized limit of the general parity-preserving curvature-squared model was considered in \cite{Nishino:2006cb} and those results were adapted in \cite{Bergshoeff:2009hq} to deduce some features of the ${\cal N}=2$
extension of the new massive gravity model. Here we present more details and give the extension to GMG; i.e. we allow for parity-violating terms.

Any ${\cal N}=2$  model can be viewed in ${\cal N}=1$ terms. In the context of the GMG models, this involves a decomposition of the ${\cal N}=2$ graviton multiplet into an ${\cal N}=1$ graviton multiplet and another ${\cal N}=1$ multiplet that propagates helicities $\pm(\frac{3}{2},1)$. We begin by presenting this new multiplet.

%%%%%%%%%%%%%%%%%%%%%%%%%%%%%%%%%%%
\subsection{The spin $(3/2,1)$ multiplet}

Consider the following infinitesimal supersymmetry transformations
connecting a `second' gravitino field $\psi'_\mu$ to a vector field
$A_\mu$ and a `second' scalar auxiliary field $S'$:
\begin{eqnarray}
\delta \psi'_\mu &=& \frac{1}{4} \gamma^\tau \gamma_\mu \epsilon A_\tau + \frac{1}{2} \gamma_\mu \epsilon S'\;, \nonumber \\
\delta A_\mu &=& \frac{1}{2}\bar\epsilon \gamma_\nu\gamma_\mu {\cal R}'{}^\nu_{(lin)} \, , \qquad
\delta S' = \frac{1}{4}\bar\epsilon \gamma \cdot {\cal R}'_{(lin)}\, .
\end{eqnarray}
It may be verified that these transformations close off-shell, up to gauge transformations, in the same way as those of (\ref{susytrans}).
The following three Lagrangians are invariant, up to a total derivative, under these transformations:
\begin{eqnarray}
L_1 &=& \bar\psi' \cdot {\cal R}'{}_{(lin)} - \frac{1}{2} A^\mu A_\mu + 2\left(S'\right)^2\;, \\
L_2 &=& \frac{1}{2} \bar\psi' \cdot {\cal C}'{}_{(lin)} - \frac{1}{4} \varepsilon^{\mu\nu\rho} A_\mu F_{\nu\rho}\;, \\
L_3 &=& -\frac{1}{2} \bar\psi'_\mu
\left(\gamma^\tau\partial_\tau\right) {\cal C}'{}^\mu_{(lin)} -
\frac{1}{4} F^{\mu\nu}F_{\mu\nu}\;,
\end{eqnarray}
where
\be
F_{\mu\nu} \equiv 2\partial_{[\mu} A_{\nu]}\, .
\ee

Putting this together we get the following Lagrangian \be L' =
L'_{(bos)} + L'_{(ferm)}\;, \ee where
\begin{eqnarray}
L_{(bos)} &=& - \frac{1}{4m^2} F^{\mu\nu}F_{\mu\nu}  -
\frac{1}{4\mu}  \varepsilon^{\mu\nu\rho} A_\mu F_{\nu\rho}  -
\frac{1}{2} A^\mu A_\mu\;,
\nonumber \\
L_{(ferm)} &=& \bar\psi'_\mu {\cal R}'{}^\mu_{(lin)} + \frac{1}{2\mu} \bar\psi'_\mu {\cal C}'{}^\mu
- \frac{1}{2m^2} \bar\psi'_\mu \left(\gamma^\nu\partial_\nu\right){\cal C}'{}^\mu_{(lin)}\, .
\end{eqnarray}
This Lagrangian propagates one helicity $(\frac{3}{2},1)$ supermultiplet with mass $m_+$ and one helicity $(-\frac{3}{2},-1)$ supermultiplet with
mass $m_-$.  In the special case that $m_-\to\infty$ for fixed $m_+$, which corresponds to the $m^2\to\infty$ limit, we have a supersymmetrization
of the  `odd-dimensional self-dual'  (or `Proca square-root') model of  \cite{Townsend:1983xs}.

%%%%%%%%%%%%%%%%%%%%%%%%%%%%%%%
\subsection{Linearized ${\cal N}=2$ massive supergravity}

The fields of the off-shell linearized ${\cal N}=2$ supergravity are
the metric perturbation $h_{\mu\nu}$, two gravitini $\psi_\mu^a$
($a=1,2$), a vector $A_\mu$ and an auxiliary scalar field $S^{ab}$
that is symmetric and traceless in its two indices, which we can
interpret as indices of the $SO(2)$ automorphism group of the ${\cal
N}=2$ supersymmetry algebra.   The ${\cal N}=2$ infinitesimal
supersymmetry transformations of these fields, with anticommuting
Majorana spinor parameters $\epsilon^a$, are
\begin{eqnarray}
\delta h_{\mu\nu} &=& \bar\epsilon^a \gamma^{}_{( \mu} \psi_{\nu )}^a \;,\nonumber \\
\delta\psi^a_\mu &=& - \frac{1}{4} \gamma^{\rho\sigma} \partial_\rho h_{\mu\sigma} \epsilon^a -
\frac{1}{4} \varepsilon^{ab}\gamma^\tau\gamma_\mu \epsilon^b A_\tau + \frac{1}{2}\gamma_\mu \epsilon^b S^{ab}\;, \nonumber \\
\delta A_\mu &=&  \frac{1}{2}\varepsilon^{ab} \bar\epsilon^a \gamma_\nu\gamma_\mu {\cal R}_{(lin)}^{\nu\, b}\;, \nonumber \\
\delta S^{ab} &=& \frac{1}{2} \bar\epsilon^a \gamma\cdot {\cal
R}_{(lin)} ^b - \frac{1}{4} \delta^{ab} \bar\epsilon^c \gamma\cdot
{\cal R}_{(lin)}^c\;.
\end{eqnarray}
The following three Lagrangians are invariant under these
transformations
\begin{eqnarray}\label{threelags}
L_1^{{\cal N}=2} &=& \frac{1}{2} h^{\mu\nu} G_{\mu\nu}^{(lin)} + \bar\psi^a \cdot {\cal R}_{(lin)}^a+ S^{ab} S_{ab} -
\frac{1}{2} A^\mu A_\mu \;,\nonumber \\
L_2^{{\cal N}=2} &=& \frac{1}{2} h^{\mu\nu} C_{\mu\nu}^{(lin)}  + \frac{1}{2} \bar\psi^a \cdot {\cal C}_{(lin)}^a
- \frac{1}{4} \varepsilon^{\mu\nu\rho} A_\mu F_{\nu\rho}\;, \nonumber \\
L_3^{{\cal N}=2} &=& - \frac{1}{2} \varepsilon^{\mu\tau\rho}
h_\mu{}^\nu \partial_\tau C_{\rho\nu}^{(lin)} - \frac{1}{2}
\bar\psi^a_\mu \left(\gamma^\tau\partial_\tau\right) {\cal
C}_{(lin)}^{\mu\, a} - \frac{1}{4} F^{\mu\nu}F_{\mu\nu}\;.
\end{eqnarray}

Putting these results together we get the following Lagrangian for
the ${\cal N}=2$ supersymmetric extension of linearized GMG: \be
L_{GMG}^{{\cal N}=2} = L_{(grav)}^{{\cal N}=2} + L_{(ferm)}^{{\cal
N}=2} + L_{(vec)}^{{\cal N}=2} + L_{(aux)}^{{\cal N}=2}\, , \ee
where
\begin{eqnarray}
 L_{(grav)}^{{\cal N}=2} &=&  \frac{1}{2} h^{\mu\nu} G_{\mu\nu}^{(lin)} +  \frac{1}{2\mu} h^{\mu\nu} C_{\mu\nu}^{(lin)}
 - \frac{1}{2m^2} \varepsilon^{\mu\tau\rho} h_\mu{}^\nu \partial_\tau C_{\rho\nu}^{(lin)}\;,  \nonumber\\
 L_{(ferm)}^{{\cal N}=2} &=& \bar\psi^a \cdot {\cal R}_{(lin)}^a + \frac{1}{2\mu} \bar\psi^a \cdot {\cal C}_{(lin)}^a
 - \frac{1}{2m^2}  \bar\psi^a_\mu \left(\gamma^\tau\partial_\tau\right) {\cal C}_{(lin)}^{\mu\, a}\;, \nonumber\\
 L_{(vec)}^{{\cal N}=2} &=& - \frac{1}{2} A^\mu A_\mu - \frac{1}{4\mu} \varepsilon^{\mu\nu\rho} A_\mu F_{\nu\rho}
 - \frac{1}{4m^2} F^{\mu\nu}F_{\mu\nu}\;, \nonumber\\
 L_{(aux)}^{{\cal N}=2} &=& S^{ab} S_{ab}\;.
 \end{eqnarray}
 These formulae show that ${\cal N}=2$ supersymmetry concisely
combines the different mechanisms in 3D of assigning mass to modes
of spin 1, $\ft{3}{2}$ and 2.

%%%%%%%%%%%%%%%%%%%%%%%%%%%%%%%%%%%%%%%%%%%%%%%%%%%%%%%%%

\section{Conclusions and Outlook}

Motivated by recent work on massive gravity theories in three
dimensions, we have constructed the full off-shell supersymmetric
${\cal N}=1$ 3D supergravity theory with cosmological and
Lorentz-Chern-Simons terms, and general curvature squared terms. The
general model of this type is parametrized by four mass parameters
$(M, \mu,m, \tilde m)$ and a dimensionless coefficient  $\sigma$ of
the Einstein-Hilbert term that is unity for standard 3D General
Relativity. We have found that the maximally symmetric vacua,
with cosmological constant $\Lambda$,  are characterized by two
curves in the $(\Lambda, M^2)$ plane, and all vacua on one of them
are supersymmetric. This family of supersymmetric vacua includes the  Minkowski
vacuum as a limiting case. Apart from this Minkowski vacuum, the overall picture is
remarkably different from that found in \cite{Bergshoeff:2009aq} for
the non-supersymmetric ``new massive gravity'' (NMG) model. This is due to
the new `auxiliary' field in the supergravity theory; although it
really is auxiliary in the NMG case, its equation of  motion is
cubic with coefficients that depend on the scalar curvature $R$.
Because of this, it is unclear whether any of the  conclusions of
\cite{Bergshoeff:2009aq} concerning unitarity in adS vacua, and the
central charges of the boundary CFTs, will still apply in the
supergravity case. Thus, one obvious direction for further research
is a unitarity/stability analysis for adS vacua.

In the context of  a possible adS/CFT relation, a crucial role is played by
the central charges of the asymptotic Virasoro algebra. While in
this paper we did not attempt to compute these charges from first
principles (as could be done, e.g., by following the original Brown-Henneaux
argument \cite{Brown:1986nw}) a natural conjecture emerges from an
application of  a formula of \cite{Saida:1999ec},  and of \cite{Kraus:2005vz}, who
have demonstrated its  applicability for generic (parity-preserving)
higher-curvature Lagrangians ${\cal L}_{3}$ with $adS_{3}$ vacuum. This formula is
 \bea
  c \ = \ \frac{\ell}{2G_{3}}g_{\mu\nu}\frac{\partial {\cal
  L}_{3}}{\partial R_{\mu\nu}}\;,
 \eea
where $G_{3}$ is Newton's constant determined by $\kappa^2=16\pi
G_{3}$. It is not clear to us whether this formula is still
applicable in our case, in which there are also terms that couple
curvature-squared terms to the  extra scalar $S$. Nevertheless, if we assume that it
is applicable, at least for the  supersymmetric adS vacua with $S^2=-\Lambda$, then we deduce that
 \bea\label{cencharge}
  c_{L} \ = \ \frac{3\ell}{2G_{3}}\left(\sigma+\frac{1}{\mu\ell}\right)\;, \qquad
  c_{R} \ = \ \frac{3\ell}{2G_{3}}\left(\sigma -\frac{1}{\mu\ell}\right)\;,
 \eea
where we have also included the known contribution of the
parity-violating Lorentz-Chern-Simons term \cite{Kraus:2005zm}. We
note, in particular,  that the values of the central charges
coincide with those of pure TMG; in other words, the extra
contributions due to the curvature squared terms (as given in
\cite{Bergshoeff:2009aq}) are precisely canceled by the new
contributions from the curvature couplings to $S$. The conjecture
that (\ref{cencharge}) indeed represents the correct central charges
is confirmed by the observation that at the chiral point
$\mu\ell=-\sigma$, at which $c_{L}=0$, the pp-wave solution
(\ref{pp}) is pure gauge (since its exponent (\ref{ppn}) becomes
$n=2$), being replaced by a `logarithmic mode' as happens in chiral
gravity (see, e.g., \cite{Maloney:2009ck}). We leave a systematic
analysis of the adS/CFT relation for the case of the massive
supergravity models given here, e.g.~along the lines of a similar
analysis for TMG \cite{Skenderis:2009nt}, to future work.

Apart from identifying the maximally supersymmetric adS vacua, we
have found the general 3D supergravity field configuration that
preserves only 1/2 of the supersymmetry.  As a Majorana 3D spinor
has just two real components, the only possible fraction less than
$1$ is $1/2$. For constant $S$ these configurations  are of pp-wave
type. Specific configurations of pp-wave type have previously been
shown to solve the equations of motion of both super-TMG and  the
bosonic NMG.   We have found the supersymmetric  pp-wave solutions
of the generic 3D supergravity within the class of theories
considered here, which differ from those of the purely gravitational
theory as a consequence of the non-linear interactions of the
supergravity scalar `auxiliary' field $S$.

A crucial issue is unitarity, and here we have presented a complete
analysis for the  linear supergravity theories obtained by
linearization about the supersymmetric Minkowski vacuum. We have
confirmed the unitarity of bosonic models previously known to be
unitary, such as NMG or its parity-violating extension to GMG, and
we have extended these results to the fermionic sector. In addition,
we have found a new unitary linearized supergravity model that
combines the LCS term of TMG with the curvature squared term of NMG.
This model propagates a single $(+2,+3/2)$ helicity multiplet, just
like super-TMG. For this reason, we have called it ``new
topologically massive gravity'' (NTMG).  However, it is currently
unclear whether this linearized theory is still consistent when
interactions are included because the interactions break an accidental 
gauge invariance of the linearized theory.

We have also constructed the linearized ${\cal N}=2$ massive
supergravity, which propagates both a multiplet of helicities
$(2,\frac{3}{2}, 1)$ and a multiplet of helicities
$(-2,-\frac{3}{2}, -1)$, in general with different masses $m_\pm$.
This model unifies the GMG model of \cite{Bergshoeff:2009hq} with
the general spin 1 theory; i.e.~the 3D Proca theory with a CS term.
In particular the spin 1 sector of the ${\cal N}=2$ super TMG is the
self-dual spin 1 model of \cite{Townsend:1983xs} whereas the spin 1
sector of the ${\cal N}=2$ super NTMG is the  topologically massive
spin 1 theory of \cite{Deser:1981wh}.  For parity preserving models
the representation theory of the super-Poincar\'e group is
essentially the same for massive 3D particles as it is for massless
4D particles, so we expect that there is an ${\cal N}=8$ massive
supergravity theory and that ${\cal N}=8$ is maximal. For parity
violating models the maximal value of ${\cal N}$ must be less than
this.

An obvious next step is the construction of the full ${\cal N}=2$
massive supergravity model. Given that the options for
maximally-symmetric vacua for ${\cal N}=1$ are so different from
those for ${\cal N}=0$, one might think that they would again be
different for ${\cal N}=2$. However, a cosmological term in an
${\cal N}=2$ theory could involve at most one scalar, and would
therefore break the $SO(2)$  symmetry. It therefore seems likely
that vacua for ${\cal N}=2$, and by extension for ${\cal N}>2$, are
determined by the truncation to ${\cal N}=1$. Thus, we expect the
results obtained here to survive the extension to higher ${\cal N}$.

One important motivation for our work that we have not yet mentioned is
the possibilty that some massive supergravity might be ultra-violet
finite. The situation for NMG, to take the simplest case, is unclear to us.
On the one hand it has been argued in \cite{Oda:2009ys} that NMG is super-renormalizable
(as one might expect from the known renormalizability of the
4D `$R+K$' theory \cite{Stelle:1976gc}). On the other hand, it was argued in
\cite{Deser:1981wh} that NMG is not even renormalizable, but even 
if this is true it is still likely that super-NMG will be better behaved than NMG,

Finally we would like to mention that in the context of
massive 3D Poincar\'e supersymmetry an unconventional multiplet
shortening may arise due to the possibility of non-central charges
in the superalgebra \cite{Bergshoeff:2008ta}. It would be
interesting to see whether such a mechanism can be realized for
massive supergravity models of the type considered in this paper.

\textit{Note added:} Shortly after the first version of this paper
appeared on the archives, a paper of Dalmazi and Mendonca appeared
\cite{Dalmazi:2009pm}, in which the model that we have here called
`new topologically massive gravity' was discussed. (See also
\cite{Dalmazi:2009bd}.)

\subsection*{Acknowledgments}

We acknowledge discussions with David Chow, Sadik Deger, Ali Kaya,
Yoshiaki Tanii and Ricardo Troncoso. EB and PKT are grateful for the
hospitality of the Benasque Centre for Physics, and ES thanks
Groningen University, SISSA in Trieste, Scuola Normale Superiore di
Pisa and Bogazici University in Istanbul, for hospitality, where
part of this work was done. We thank T.~Nutma for providing
assistance with the figures. PKT is supported by an EPSRC Senior
Fellowship. The research of ES is supported in part by NSF grants
PHY-0555575. The work of OH is part of the research programme of the
`Stichting voor Fundamenteel Onderzoek der Materie (FOM)'.

\end{document}